\documentclass[9pt,twocolumn,twoside]{osajnl}
\usepackage{multirow}
\usepackage{subcaption}  % amsmath
\usepackage{amssymb,amsmath}
\journal{ao} % Choose journal (ao, aop, josaa, josab, ol)

\setboolean{shortarticle}{false} 
\title{Robust 3D Surface Recovery by Applying a Focus Criterion in White Light Scanning Interference Microscopy}

\author[1]{Hernando Altamar-Mercado}
\author[1]{Alberto Patiño-Vanegas}
\author[2*]{Andres G. Marrugo}

\affil[1]{Facultad de Ciencias Basicas, Universidad Tecnologica de Bolivar, Cartagena, Colombia.}
\affil[2]{Facultad de Ingenieria, Universidad Tecnologica de Bolivar, Cartagena, Colombia.}
\affil[*]{Corresponding author: agmarrugo@utb.edu.co}

\ociscodes{}
\begin{abstract}
White Light Scanning Interference (WLSI) microscopes provide an accurate surface topography of engineered surfaces. However, the measurement accuracy is substantially reduced in surfaces with low reflectivity regions or high roughness, like a surface affected by corrosion. An alternative technique called Shape from Focus (SFF) takes advantage of the surface texture to recover the 3D surface by using a focus metric through a vertical scan. In this work, we propose a technique called SFF-WLSI, which consists in recovering the 3D surface of an object by applying the Tennegrad Variance (TENV) focus metric to WLSI images. Extensive simulation results show that the proposed technique yields accurate measurements under different surface roughness and surface reflectivity outperforming the conventional WLSI and the SFF techniques. We validated the simulation results on two real objects with a Mirau-type microscope. The first, a flat lapping specimen with $R_a=0.05~\mu\mathrm{m}$ for which we measured an average value of $R_a=0.055~\mu\mathrm{m}$ and standard deviation $\sigma=0.008~\mu\mathrm{m}$. The second, a metallic sphere with corrosion which we reconstructed with WLSI versus the proposed SFF-WLSI technique producing a better 3D reconstruction with less undefined depth values.
\end{abstract}

\setboolean{displaycopyright}{false}

\begin{document}

\maketitle

\section{Introduction}

White Light Scanning Interference (WLSI) microscopes provide unbeatable surface topography measurements of engineered surfaces~\cite{harding2013handbook}. They deliver vertical resolution down to a fraction of a nanometer while maintaining the submicron lateral resolution measurements found in any typical microscope. 
The surface height is measured by detecting the best focus position
in the interference signal observed by a single pixel during an axial scan~\cite{Gianto2016,Larkin1996,Montgomery2013}.
Interference signals from rough surfaces (above $R_a = 50$~nm) are analyzed by identifying the intensity maximum position or the peak envelope position, and not with phase techniques because the phase information loses its meaning in rough surfaces~\cite{1992ApOpt..31..919D}.
However, when measuring surfaces with regions of low reflectivity or varying degrees of roughness, like a surface affected by corrosion, the detection of the interference pattern at these regions is substantially reduced. This shortcoming of WLSI microscopes is known~\cite{Gao:2007gz}, and it has to be considered before using WLSI for measuring such surfaces.

As an alternative, microscopic Shape from Focus (SFF)~\cite{Noguchi:1996gn} is another technology that exploits the limited Depth of Field (DOF) of microscopes as a way to recover the 3D shape of an object by the estimation of the best focus position through an axial $z$-scan. There are many implementations of SFF~\cite{Florczak2014,Helmli2001,Tian2017}. However, most methods are designed around a focus quality metric that quantifies the difference between a sharp local region of an image and an out-of-focus one. Various focus metrics have been proposed including spatial, entropy, and frequency-based metrics. All of which have their advantages and disadvantages, but their performance is dependent on the object properties~\cite{Pertuz:2013gha}.

In contrast to WLSI microscopes, SFF microscopes take advantage of varying surface roughness and changes in reflectivity that give rise to images with great amount of texture. As a result, the focus metric yields a single maximum which enables the detection of the focus position. Although, due to noise and textureless regions, the focus metric often exhibits local maxima which hinders the accurate localization of the focus quality peak along the range of interest~\cite{Xu2014}. This drawback is often overcome by projecting texture on the imaged surface~\cite{Lee2013,Noguchi1996}. Moreover, while in optimal conditions the vertical resolution of a WLSI system can be up to a fraction of a nanometer, in SFF the vertical resolution is determined by the DOF of the optical system which can be as large as 100 $\mu$m for $5\times$ magnification and can only go down to a few microns for the highest magnifications.

To overcome the disadvantages of the SFF systems, several authors have proposed sophisticated strategies, like using a 3D template instead of a 2D template for processing the stack of focus images~\cite{Fan:2018dd}. Although the performance of the system does improve, the computational cost increases and the vertical resolution is the same as any other SFF system. Nevertheless, there have been previous works which include focus measures in interferometric setups for in-focus plane detection~\cite{Lizewski:2014ez}, for vibration detection~\cite{Filipinas:2012ec}, among other applications. 
Based on this idea, we propose a method that consists in applying a focus criterion in WLSI microscopy to achieve the best performance for surfaces with varying roughness and reflectivity. Our contributions are three-fold. First, we propose an approach for achieving WLSI vertical resolution based on an SFF-operated low-magnification microscope. Second, by applying SFF on WLSI microscopy images, we obtain a robust profilometer that performs well in textureless and low-reflectivity regions. Third, we propose a simulation approach to test the performance of the 3D surface recovery method with objects of different roughness and its image acquisition by standard and WLSI microscopy. 

\section{Theory fundamentals}
\label{sec:theory}

\subsection{Two-beam Interferometer Response}

\begin{figure}[tbp]
\centering
\fbox{\includegraphics[width=0.95\linewidth]{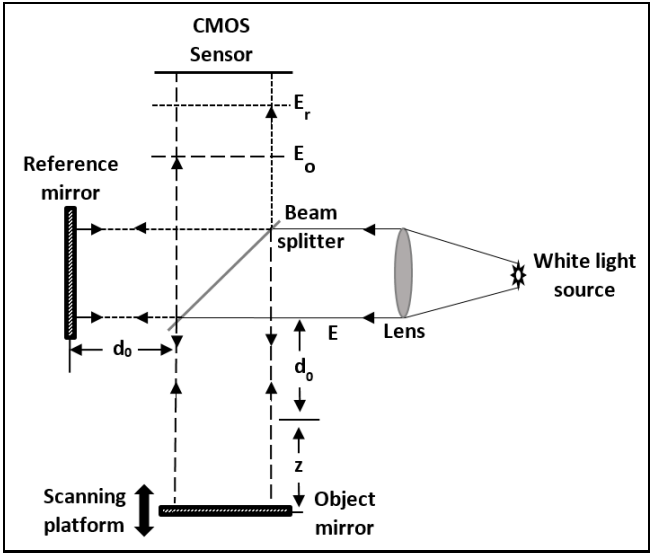}}
\caption{Schematic of a two-beam interferometer.}
\label{fig:interferometro}
\end{figure}

The intensity pattern at the position $ (x, y) $ in the output plane of a two-beam interferometer for an object mirror placed at position $z$, as shown in Fig.~\ref{eq:intensity_2beam}, is given by
\begin{align}
\begin{split}
	I_z(x,y) =& I_0(x,y) + I_r(x,y) + \\ & 2\left(I_0(x,y) I_r(x,y)\right)^{\frac{1}{2}} \gamma \left(\frac{2z}{c}\right) \cos \left[ \frac{4\pi}{\bar{\lambda}}z + \Phi_0 \right] \enspace,
\end{split}
\label{eq:intensity_2beam}
\end{align}
where $I_0$ and $I_r$ are the object and reference intensity beams associated with the electric fields $\textbf{E}_0$ and $\textbf{E}_r$, respectively. $\gamma \left(2z/c\right)$ is the visibility function of the interference fringes, $2z$ is the optical path difference (OPD) from the two beams, $c$ is the speed of light, $\bar{\lambda}$ is the mean wavelength of the light source, and $\Phi_0$ is the additional phase introduced by the components of the optical system. 

For a white light source with a Gaussian spectral density function, \eqref{eq:intensity_2beam} can be written as
\begin{align}
	I_z(x,y) = I_0(x,y)\left\lbrace 1 + V \exp\left[-\frac{4z^2}{L_c}  \right] \times \cos \left[ \frac{4\pi}{\bar{\lambda}}z + \Phi_0 \right] \right\rbrace \enspace,
\label{eq:WLI_intensity}
\end{align}
where $V$ is the visibility of the interference fringes, and $L_c=\frac{c}{\pi \Delta \nu}$ is the coherence length of the white light source with spectral width $\Delta \nu$.

\subsection{Imaging system}
\label{sec:imaging_system}
\begin{figure}[tbp]
\centering
\fbox{\includegraphics[width=0.95\linewidth]{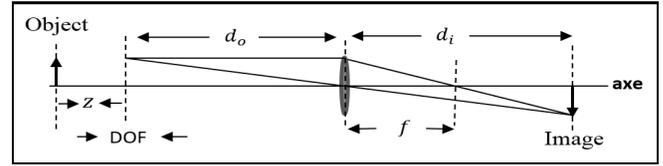}}
\caption{A thin lens imaging system.}
\label{fig:formador_imagen}
\end{figure}

The simplest imaging system consists of a thin lens of focal length $f$. If a plain object is located in front of a lens at a distance $d_o$, the lens forms a sharp image that can be registered as an intensity distribution $ I_f $ by a sensor located at a distance $d_i$ from the lens, as shown in Fig.~\ref{fig:formador_imagen}. The condition for a sharp image is given by the thin lens equation,
\begin{equation}
	\frac{1}{f} = \frac{1}{d_o} +\frac{1}{d_i} \enspace.
    \label{eq:focused_Image}
    \end{equation}   
If the object is displaced a distance $z$ from the $d_o$ position, then the recorded image $I(z)$ at the $d_i$ position is a blurry version of the image $I_f$. The image is accepted focused if the object remains within the $DOF$, for which the image remains acceptably sharp.
The recorded blurry image $I(z)$ is given by
\begin{equation}
	I(z) = I_f \ast h_z + \eta \enspace,
    \label{eq:intensity_Image}
    \end{equation}
where $\ast$ is the convolution operator, $h_z$ is the point spread function (PSF) of the imaging system and $\eta$ is the electronic noise level of the sensor.
The DOF of the system~\cite{Lim2012} can be calculated as
\begin{equation}
	DOF = \frac{\lambda n}{N_A^2} +\frac{n}{  \mathcal{M} \cdot N_A}e \enspace,
\label{eq:Dof}
\end{equation}
where $\lambda$ is the wavelength of the light source, $N_A$ the numerical aperture, $\mathcal{M}$ is the lateral magnification of the lens, $n$ is the refractive index of the medium, and $e$ is the pixel size of the camera.

\subsection{Focus Measure}
\label{sec:focus_measure}

The sharpness of an image is evaluated by focus metrics. In this work, we have used the Tennegrad Variance (TENV) metric~\cite{Pertuz:2013gha} based on the variance of the image gradient defined as 
\begin{align}
	FM_{x,y} = \sum\limits_{i,j\; \in \; \Omega \left(x,y \right)}\left[G_{i,j}-\Bar{G} \right]^2 \enspace,
    \label{eq:TENV_Metrics}
\end{align}
where $\bar{G}$ is the mean value of the magnitude of the image gradient in the window $\Omega$. The magnitude of the gradient is calculated as $G = \sqrt[]{G_x ^ 2 + G_y ^ 2}$, where $G_x$ and $G_y$ are the gradients of the image in the $x$ and $y$ directions, respectively. These are calculated by convolution of the image with the Sobel filter~\cite{SobelI.andFeldman1968}.

\subsection{3D Shape Recovery}
The 3D shape recovery in a vertical scanning microscope is usually obtained either using a focus metric (SFF) or with an interferometric objective lens by processing the interference images (WLSI). In both methods, the object is scanned vertically by acquiring a stack of images for different depth values. In SFF, the depth of each pixel is obtained by calculating the focus metric at every scan position in the stack of images, which yields a focus curve. This curve contains a peak that corresponds to the best focus position at a depth $z$. After the depth value for every pixel is obtained, the collection of depth values is the depth map or 3D shape of the object. 

Alternatively, in WLSI the interferometer superimposes fringes on the image at the sensor from points that meet the OPD condition. The depth at each pixel is obtained by processing the intensity signal recovered for every position in the stack of images. This signal is an intensity modulated curve, given by~\eqref{eq:intensity_2beam}, in which the position where the fringe visibility is maximum corresponds to the best focus position and hence the pixel depth. Although there are many methods for processing WLSI signals~\cite{Vo:2017dp}, we use here the maximum intensity detection because it is fast, requires little data storage, and has been shown to yield accurate results in surfaces of relatively high slope scanned with fine vertical stepping~\cite{harding2013handbook}.  

\section{Description of the Method}

Having described the limitations of WLSI and SFF vertical scanning microscopy systems, we describe here the proposed methodology for improving the performance of a 3D microscopic optical profilometer.

\subsection{Proposed Method}

Starting from a Mirau-type WLSI microscope, we assume that under conventional surfaces with high reflectivity the system yields accurate surface topography. However, under low-reflectivity or high-roughness regions, the system produces inaccurate results which we regard as points with an undefined depth value or spikes. Our proposed approach, called SFF-WLSI, consists of recovering the 3D surface of the object by applying the TENV focus metric to the stack of interference images as if those images were obtained by SFF vertical scanning. We argue that by processing interference images with a focus metric we obtain a robust 3D profilometer that performs: 1) as well as a conventional WLSI system for high reflectivity and smooth surfaces, and 2) much better than the typical SFF system for surfaces with high roughness and regions with low reflectivity.
Given that the coherence length of a white light source is typically smaller than the DOF of the microscope objectives commonly used in surface metrology, the SFF-WLSI method enables the precise determination of the best focus position better than SFF and at par with WLSI for high reflectivity surfaces.

\begin{figure}[t]
\centering
\fbox{\includegraphics[width=0.85\linewidth]{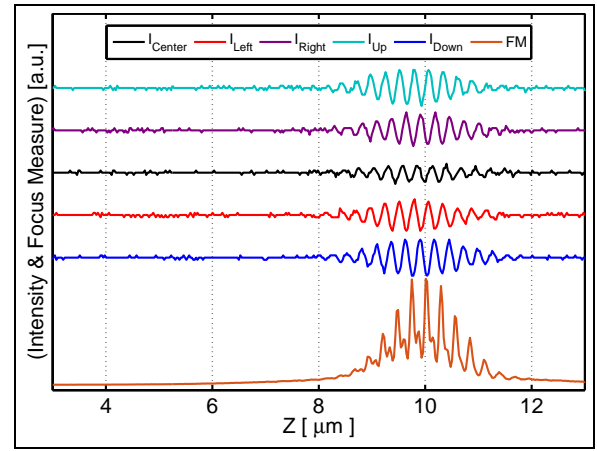}}
\caption{An example of a WLSI signal for a low reflectivity point (black line), the signals for several neighbor pixels (left, right, up, down), and the curve from applying the focus metric (FM) in a $5\times5$ window centered in the low reflectivity point.}
\label{fig:interferograms_colors}
\end{figure}

Nevertheless, our aim is to have a system that also performs well under low reflectivity surfaces. 
In conventional WLSI, a low reflectivity surface produces low signal-to-noise-ratio (SNR) signals in which precise determination of the best focus position is hindered, as shown in the $I_{\text{Center}}$ curve in Fig.~\ref{fig:interferograms_colors}. This low SNR can be explained in terms of \eqref{eq:intensity_2beam}, where the reference beam $I_r$ dominates over a low-value $I_0$ beam coming from the object surface. As a result, the fringe visibility decreases.
However, note in Fig.~\ref{fig:interferograms_colors} that there is an important relative spatial difference between the signals as each one is passing through its zero OPD. Moreover, the signals are not in phase, as each point has a different depth, and the envelopes are not exactly symmetric which impedes centroid-based focus detection.  

The proposed windowed processing of the WLSI images enables the measurement of the relative intensity change of a point with respect to its neighbors as shown in the FM curve at the bottom of  Fig.~\ref{fig:interferograms_colors}. Due to the height difference of the points within the window and the broad bandwidth of the source spectrum, during a z-scan and near the best focus position, the interference pattern introduces an abrupt space-varying intensity change within the window that is quantifiable with a gradient-based focus metric. Moreover, because this measurement is based on a relative change within a small window, it is more robust at identifying the best focus position than what is achieved with the interference signal detected at a single pixel along the z-scan.
The windowed processing of the inference images also enables robustness to vibrations and noise.

In the following sections, we describe the simulations we carried out to compare the proposed method with WLSI and SFF, and the experimental results on real objects. The performance comparison cannot be carried out in a real microscope because it implies changing objective lenses, which is not practical and requires image registration.

 \section{Simulation for Performance Analysis}
\label{sec:simulation}
To compare the proposed method to WLSI and SFF, we simulate the acquisition of images from a standard microscope and a Mirau interferometric objective microscope. Because the different 3D shape estimation methods are sensitive to surface reflectivity, especially in interferometry, we generate different reflectance maps for the same object. These maps are related to surface roughness. We use the different reflectance maps to highlight the advantages and limitations of each method.

The simulation consists in:
1)~Generating ten objects of fixed shape and variable surface roughness.
2)~Two optical imaging systems; A standard optical microscope, and an interference microscope.
3)~The vertical scanning acquisition, which consists in acquiring one image for $N$ object positions over the $z$ axis. 
4)~Recovering the 3D shape by SFF for the standard microscopy images, by maximum intensity criterion for the WLSI images, and by the proposed method SFF-WLSI on the WLSI images.

\subsection{Simulation of the object}
%\label{sec:objects_simulation}

In general, the surface of an object can be difficult to describe mathematically, but it can be modeled by a function that corresponds to the shape that the manufacturer wants to print in addition to another part associated with the tool used in the manufacturing process. The second part is associated with the roughness of the surface.
We will model the surface $S$ of the object as $S = \mathcal{W} + R$, where $\mathcal{W}$ is a function of the shape and $R$ a function of roughness. We will neglect the effects associated with defects in the manufacturing machine tool that introduce surface undulations known as waviness.

\subsubsection{Object shape}
The fixed shape for the objects is a Gaussian cap with a height of 25~$\mu$m calculated with a 2D Gaussian function given by
$ \mathcal{W} = A \exp \left[-\alpha \left( x^2 + y^2  \right) \right]$, with $A=25\;\mu$m and $\alpha = 5.0 \times 10^{-6} \space {\mu m}^{-2}$.

\subsubsection{Surface Roughness}
The roughness is modeled by using the Weierstrass-Mandelbrot~(W-M) multivariate function developed by Ausloos and Berman~\cite{ausloos1985multivariate} and later used in three-dimensional elastoplastic models~\cite{yan1998contact}, and surface roughness characterization and modeling. The multivariate W-M function in Cartesian coordinates is expressed as 
\begin{equation}
\begin{split}
	R(x,y) =& C\sum\limits_{m=1}^M\sum\limits_{n=-\infty}^\infty \gamma^{(D_s-3)n} \times  \Bigg \{ \cos\Phi_{m,n} - \\ & \cos \left[ \frac{2\pi \gamma^n (x^2+y^2)^{\frac{1}{2}}}{\mathcal{L}} \cos\left( \arctan \left(\frac{y}{x} \right) - \frac{\pi m}{M} \right) + \Phi_{m,n} \right] \Bigg \},
\end{split}
\label{eq:reflectivity}
\end{equation}
where $C = \mathcal{L}(\frac{K}{\mathcal{L}})^{D_s-2} \left ( \frac{\ln \gamma}{M} \right )^\frac{1}{2}$, $D_s$ is the fractal dimension, $2 <D_s <3$, $\gamma> 1$ is a parameter that governs the frequency and heights of the consecutive cosine shapes and thus controls the spatial frequency density of the profile curve of the roughness function $R$, $K$ is a constant that acts as a scale factor, $\mathcal{L}$ the maximum extension of the surface to be simulated, M is the number of modes for the roughness simulation, and $\Phi_{m,n}$ is a random phase in the range $[0,2\pi]$.
The Weierstrass-Mandelbrot multivariate function has been truncated to the integer given by
\begin{equation}
	n_{max} = \left \lceil \frac{\ln \left( \frac{\mathcal{L}_{max} }{\mathcal{L}_{min} } \right)}{\log \gamma} \right \rceil \enspace.
    \label{eq:n_Max}
    \end{equation}
In this work, the maximum dimension observed of the object was taken as $\mathcal{L}_{\max}= 355.52\; \mu$m, and $\mathcal{L}_{\min} = 0.7222\;\mu$m as the limit of the transverse resolution imposed by the diffraction for a $20 \times$ microscope objective with numerical aperture $NA = 0.45$ using a light source with a mean wavelength of $630$~nm.

Ten roughness functions were calculated by varying the fractal dimension $D_s$ in the range of $2.10$ to $2.55$ leaving $\gamma = 1.6$ fixed.
The mean arithmetic roughness parameter $R_a$ for each simulated object was calculated in a grid of $481\times 641$ pixels as
\begin{equation}
	R_a = \frac{1}{L_x L_y} \sum\limits_{i=1}^{L_x L_y} | R_i | \enspace.
    \label{eq:Rougs_arit}
    \end{equation}
\subsubsection{Reflectance Maps}
\begin{figure}[t]
\centering
\fbox{\includegraphics[width=0.9\linewidth]{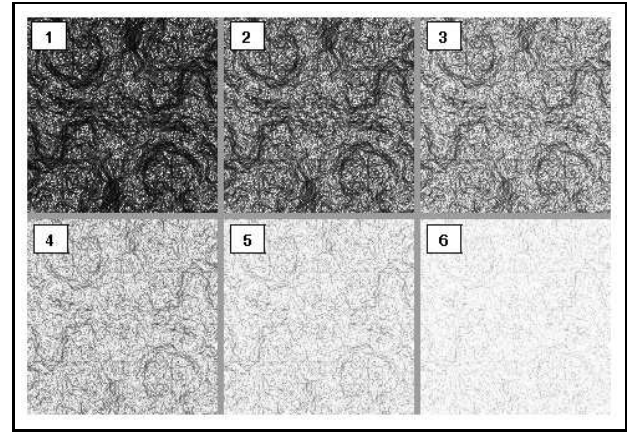}}
\caption{Reflectance maps of six of the ten simulated objects. Setting $\gamma = 1.6$ for all objects and varying the roughness from $R_a = 49.58$~nm for object 1 to $R_a=42.08$~nm for object 6. The roughness values for all objects are in Table~\ref{tab:RMSE_table}.}
\label{fig:Reflectivity Maps}
\end{figure}

The surface geometry and roughness establish a value of reflectance that plays an important role when observing objects through a microscope.
Accurate reflectance numerical modeling is still a challenging problem, especially in computer graphics for generating realistic images of scenes or for predicting image degradation in optical mirror surfaces, because it requires computational intensive methods like the Beckmann-Kirchhoff theory for rough surface reflectance or the generalized Harvey-Shack surface scatter theory~\cite{Ragheb:2003if,2012ApOpt..51..535C}. However, the purpose of our simulations is to evaluate the performance of different 3D reconstruction methods under surfaces of variable reflectance. Therefore, we used a simple model based on Lambert's cosine law for reflection as a means to determine a variable reflectance map for each object, i.e.,
\begin{equation}
	I_0 = I_i\cos \theta \enspace,
    \label{eq:Lamb_law}
    \end{equation}
where $I_0$ is the reflected intensity by the surface of the object, $I_i$ the incident intensity and $\theta$ is the angle between the normal direction at each point of the surface and the direction of observation. In the appendix, the equation for $\cos \theta$ is deduced, and we have taken $I_i = 1$.
In Fig.~\ref{fig:Reflectivity Maps} we show six of the ten reflectance maps of the modeled objects.
%==================================================================
\subsection{Simulation of a Standard Optical Microscope and an Interference Microscope}
\label{sec:Systems_simulation}
%===============================================================
The simulation of these two systems is necessary to be able to compare the SFF, WLSI and SFF-WLSI techniques with the same conditions of magnification and alignment.
Interferometric optical profilers are based on standard microscopes where an interferometer built into the objective replaces the standard objective. The interference signal obtained with these objectives is analyzed to provide quantitative data about a measured object~\cite{malacara2007optical}. 

\begin{figure}[t]
\centering
\fbox{\includegraphics[width=0.9\linewidth]{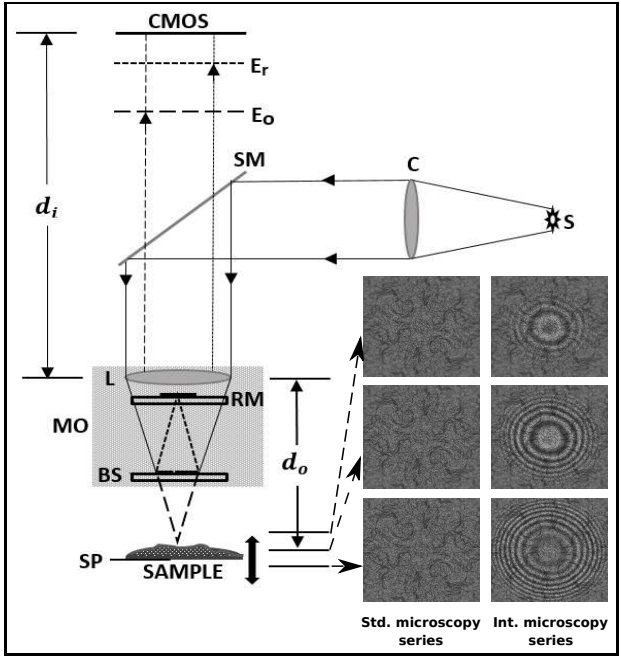}}
\caption{Schematic of a Mirau objective interference microscope.}
\label{fig:mirau_microscope}
\end{figure}

In our simulation of the optical system, we assume a microscope with a Mirau objective interferometer~\cite{kino1996confocal}, as shown in Fig.~\ref{fig:mirau_microscope}. The Mirau microscope has a beam-splitter (BS) below the objective lens (MO), and a reference mirror (RM) placed underneath a long working distance objective. At the beam-splitter, approximately half of the incident light from the objective lens is reflected toward the reference mirror, the remainder is transmitted to the sample. After reflection from the sample and the reference mirror, the two light beams $E_r$ and $E_o$ recombine at the beam-splitter and pass back through the objective lens to the CMOS detector. To obtain fringes at the best focus position, the reference mirror is set at the best focus of the objective to obtain the zero OPD~\cite{malacara2007optical}. The system calculates the surface height at each detector pixel by vertical scanning and acquiring a number of frames. During the measurement, the interference signal varies by changing the OPD between the object and reference beams. 
The values used for the simulation were: $f=4.4$~mm, $d_0=4.42$~mm, $N_A=0.45$ for a $20~\times$ microscope objective, and a camera sensor parameter $k=1.6 \cdot 10^3~\text{pix/mm}$.
For all experiments the acquisition noise in \eqref{eq:intensity_Image} was modeled as Gaussian with zero mean and $\sigma ^2=5.7\times10^{-4}$.

\subsubsection{Simulation of Standard Microscopy Images}
\label{sec:Simul_Blur}

We simulate the image $I(z)$ acquired by the microscope following the model given by~\eqref{eq:intensity_Image}, where $I_f$ comes from the light reflected from the surface of the object and is given by $I_0$ in~\eqref{eq:Lamb_law}. We assume the convolutional model with a Gaussian PSF
\begin{equation}
	h_z(x,y) = \frac{1}{{2\pi\sigma}^2}  \exp \left( -\frac{x^2+y^2}{2\sigma} \right) \enspace.
    \label{eq:PSF_Func}
    \end{equation}
This model is accepted under the limits of diffraction if one works with incoherent illumination~\cite{Subbarao1993}.
The parameter $\sigma$ sets the degree of blur at each point in the image and depends on the z-axis position, the lens, and the sensor parameters. Pentland~\cite{Pentland1987} derived an expression for $\sigma$ given by
 \begin{equation}
	\sigma(x,y,z) = 2kN_{A} f^2 \frac{| z-d_0 |}{z\left(d_0-f \right)}\enspace,
    \label{eq:blur_param}
    \end{equation}
where $N_A$ and $f$ are the numerical aperture and focal length of the lens respectively, $k$ is a constant that depends on the array of sensors and $d_0$ is the perfect focus position.
The simulated standard microscopy image is shown in Fig.~\ref{fig:BlurImage_1}. We highlight a region of low reflectivity with a white dashed-line circle and a region of high reflectivity with a white solid-line circle.

\subsubsection{Simulation of Interferometric Microscopy Images}
\label{sec:Simul_Interf}

\begin{figure}[t]
\centering
\begin{subfigure}[b]{0.18\textwidth}
\captionsetup{justification=centering}
\fbox{\includegraphics[width=\textwidth]{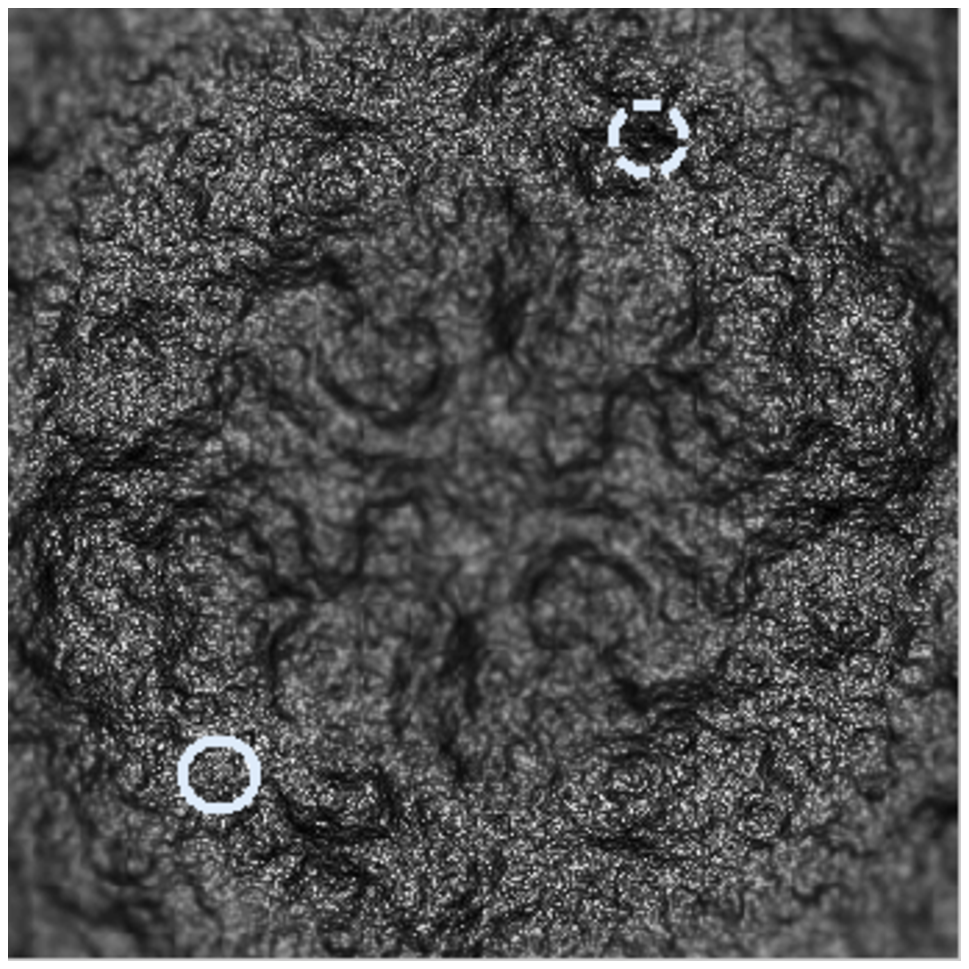}}
\caption{ }
\label{fig:BlurImage_1}
\end{subfigure}
\qquad%  ~ %add desired spacing between images, e. g. ~, \quad, \qquad, \hfill etc. 
    \begin{subfigure}[b]{0.18\textwidth}
    \captionsetup{justification=centering}
    \fbox{\includegraphics[width=\textwidth]{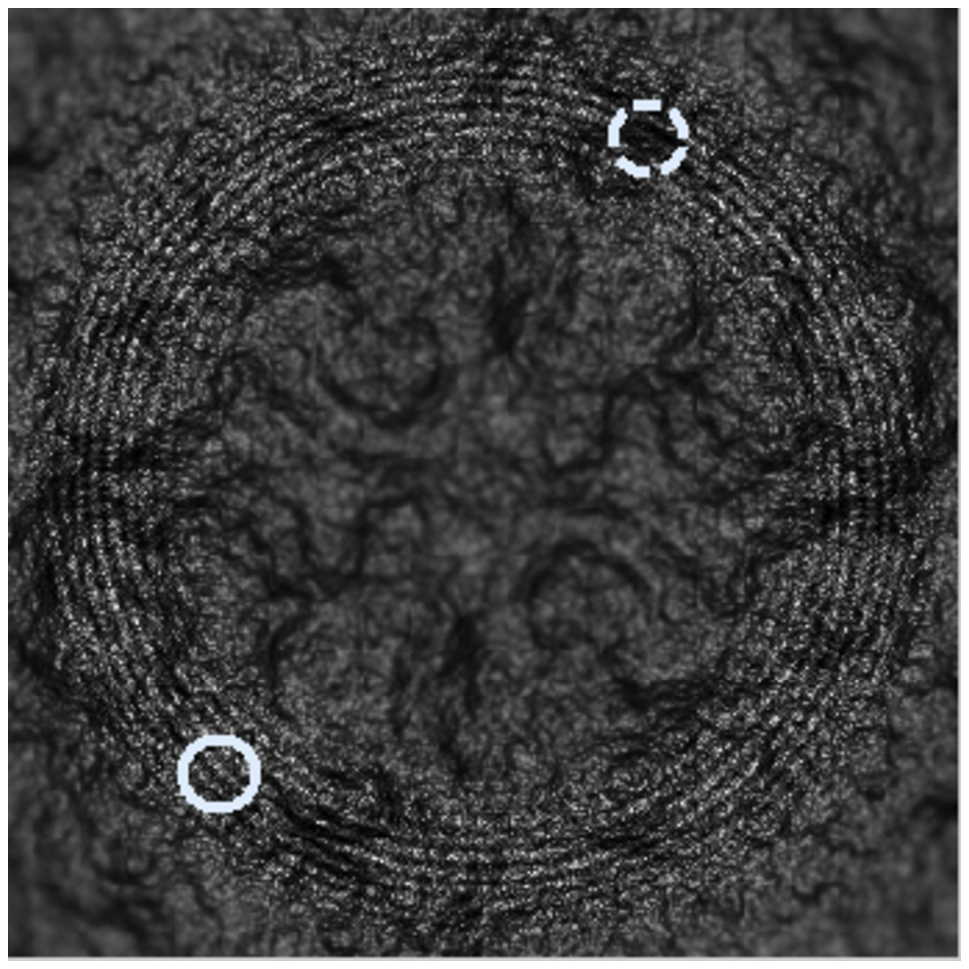}}
        \caption{ }
 \label{fig:IntImage1}
 \end{subfigure}
%\qquad%  ~ %add desired spacing between images, e. g. ~, \quad, \qquad, \hfill etc. 
    \begin{subfigure}[b]{0.18\textwidth}
    \captionsetup{justification=centering}
    \fbox{\includegraphics[width=\textwidth]{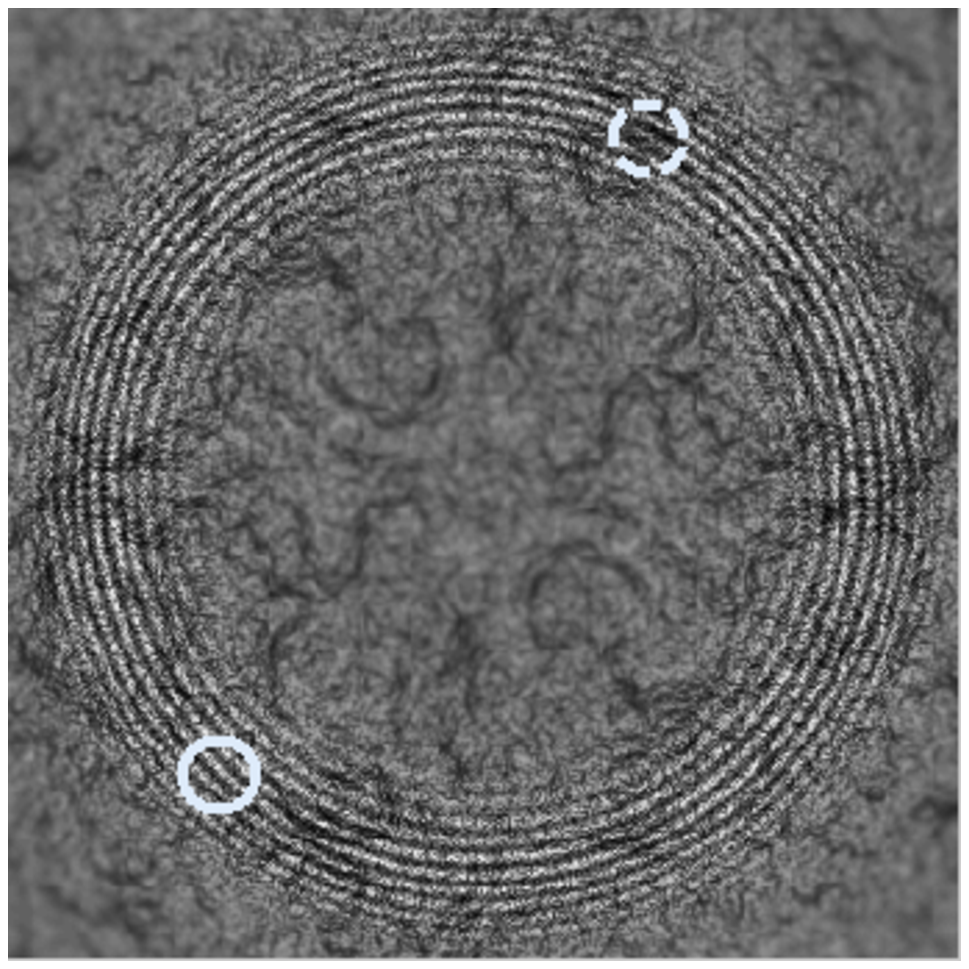}}
 \caption{ }
 \label{fig:IntImage2}
    \end{subfigure}
    \qquad
    \begin{subfigure}[b]{0.18\textwidth}
    \captionsetup{justification=centering}
    \fbox{\includegraphics[width=\textwidth]{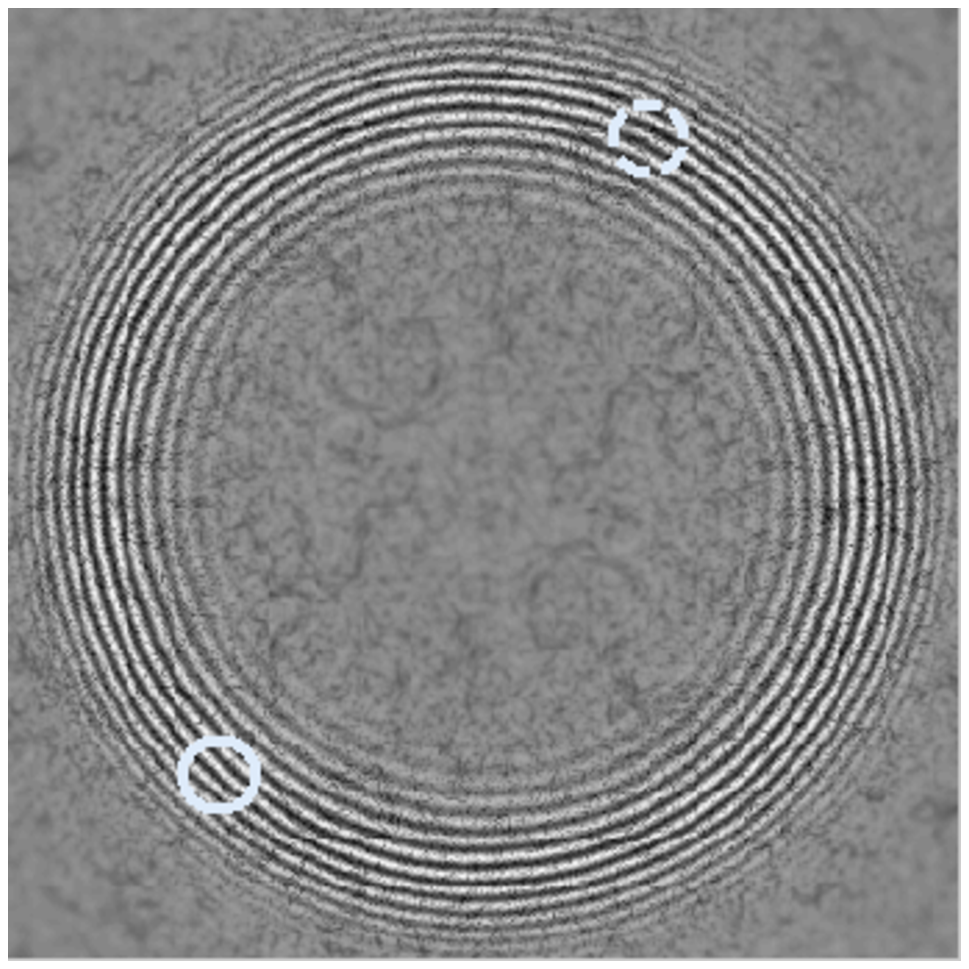}}
    \caption{ }
\label{fig:IntImage3}
\end{subfigure}
\caption{Synthetic microscopy images for Gaussian cap shaped objects with different roughness $R_a$. The solid-line white circle is a region of high reflectance; the dashed line circle is a region of low reflectance. (a) standard microscopy image for an object with $R_a = 49.58$~nm, (b)-(d) WLSI images for objects with $R_a$ = [49.58~nm, 47.00~nm, 43.82~nm].}
\label{fig:Interferograms}
\end{figure}

The image $I(z)$ produced by an interferometric microscope was simulated with~\eqref{eq:intensity_Image} as was done in the previous subsection, but $I_f$ is now the result of~\eqref{eq:WLI_intensity} with $ I_0 $ being the reflected light from the surface of the object given by~\eqref{eq:Lamb_law}.
The simulated interference microscopy images are shown in Figures~\ref{fig:IntImage1}-\ref{fig:IntImage3}. The regions of low and high reflectivity are highlighted in them, as was done in Fig.~\ref{fig:BlurImage_1}. All interference patterns were calculated with a fringe visibility $V = 0.4$. 

\subsection{Simulation of Vertical Scanning}

We simulated the vertical scanning acquisition of images of both standard and interference microscopy by moving axially at 26~nm steps. For each of the ten reflectance maps, we calculated 1000 interference patterns (one per blur step) so that each point of the object is in perfect focus in at least one of the images.

\section{Simulated results and Discussion}
\label{sec:experiments}

\subsection{Performance of the Focus Metric in Standard Microscopy and Interference Microscopy}
\label{sec:Simmul_result}

To verify the improvement in axial resolution when using a focus metric in WLSI images with respect to SFF in standard microscopy, we take a point on the object with $Z_{\text{ref}}=5.0656\; \mu$m and reflectance $R = 0.5$, with $R_a = 43.82$~nm.
This point has an intermediate value between the highest reflectance ($R = 1.0$) and the lowest reflectance ($R = 0.0$), the roughness value is also an intermediate value between the roughness values shown in Table~\ref{tab:RMSE_table}. 
At this point, the value of the focus metric was calculated for each $Z$ position using~\eqref{eq:TENV_Metrics} with a window $\Omega$ of $5\times 5$ pixels, both in standard microscopy images (SFF) and in WLSI images. 

\begin{figure}[t]
\centering
\fbox{\includegraphics[width=0.8\linewidth]{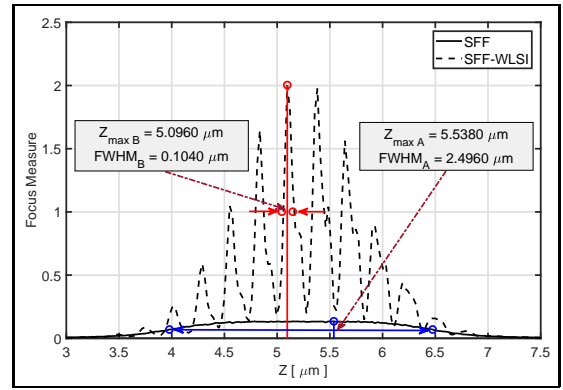}}
\caption{Value of the TENV focus metric for each Z position at a an object point with reflectance $R=0.5$ using: standard optical microscopy SFF (solid line) and SFF-WLSI with fringe visibility $V~=~0.4$ (dashed line).}
\label{fig:WLSI_FM_curves}
\end{figure}

In SFF, the window size selection comes from a trade-off between spatial resolution and robustness to the lack of texture. Large windows tend to reduce the quality of the reconstruction by excessively smoothing the depth-map, whereas small windows yield unreliable measurements in uniform regions~\cite{Pertuz:2013gha}. In this work, we found the 5x5 window to yield the best trade-off. 
The results are shown in Fig.~\ref{fig:WLSI_FM_curves}.
First, the solid line is the response from the TENV metric applied to standard microscopy images (SFF). Note that there is a range of values on the $ Z $ axis for which the focus metric curve presents a plateau shape without a well-localized maximum. For the reference point taken, the maximum of the focus metric is located at $Z_{\max A} = 5.5380~\mu $m with an axial full-width at half-maximum (FWHM) $\text{FWHM}_A = 2.496~\mu$m. 
Second, the dashed line is the response from the TENV metric applied to interferometric microscopy images (SFF-WLSI). The curve presents a global maximum in a position that can be determined without difficulty. For this curve, $ Z_{\max B} = 5.096~\mu$m with $ \text{FWHM}_B = 0.104~\mu$m, which represents a 24-fold increase in axial resolution.

\begin{table}[t]
\centering
\caption{\bf RMSE 3D reconstruction results for SFF in standard microscopy, WLSI and the proposed SFF-WLSI.}
\label{tab:RMSE_table}
\begin{tabular}{c c c c c c}
\hline
\multirow{2}{*}{Maps} & \multirow{2}{*}{$D_s$} & \multirow{2}{*}{$R_a$ \space [nm]} & \multicolumn{3}{c}{RMSE \space[$\mu$m] }   \\ \cline{4-6} 
                         &                     &                          & SFF    & WLSI   & SFF-WLSI \\ \hline
1  & 2.10 & 49.58   & 0.0859 & 4.6561 & 0.0303    \\ 
2  & 2.15 & 48.42   & 0.1006 & 3.5892 & 0.0301    \\ 
3  & 2.20 & 47.00   & 0.1185 & 2.3080 & 0.0297    \\ 
4  & 2.25 & 45.44   & 0.0963 & 0.9497 & 0.0300    \\ 
5  & 2.30 & 43.82   & 0.0893 & 0.1110 & 0.0294    \\ 
6  & 2.35 & 42.08   & 0.1035 & 0.0284 & 0.0283    \\ 
7  & 2.40 & 40.23   & 0.0796 & 0.0293 & 0.0280    \\ 
8  & 2.45 & 38.37   & 0.1207 & 0.0292 & 0.0285    \\ 
9  & 2.50 & 36.43   & 0.3602 & 0.0291 & 0.0287    \\ 
10 & 2.55 & 34.47   & 2.7817 & 0.0291 & 0.0289    \\  \hline
\end{tabular}
\end{table}

\subsection{Analysis in Points with Low and High Reflectance}
\label{sec:Des_lowReflectance}

\begin{figure*}[htbp]
\centering
\begin{subfigure}[b]{0.3\textwidth}
\captionsetup{justification=centering}
\fbox{\includegraphics[width=\textwidth]{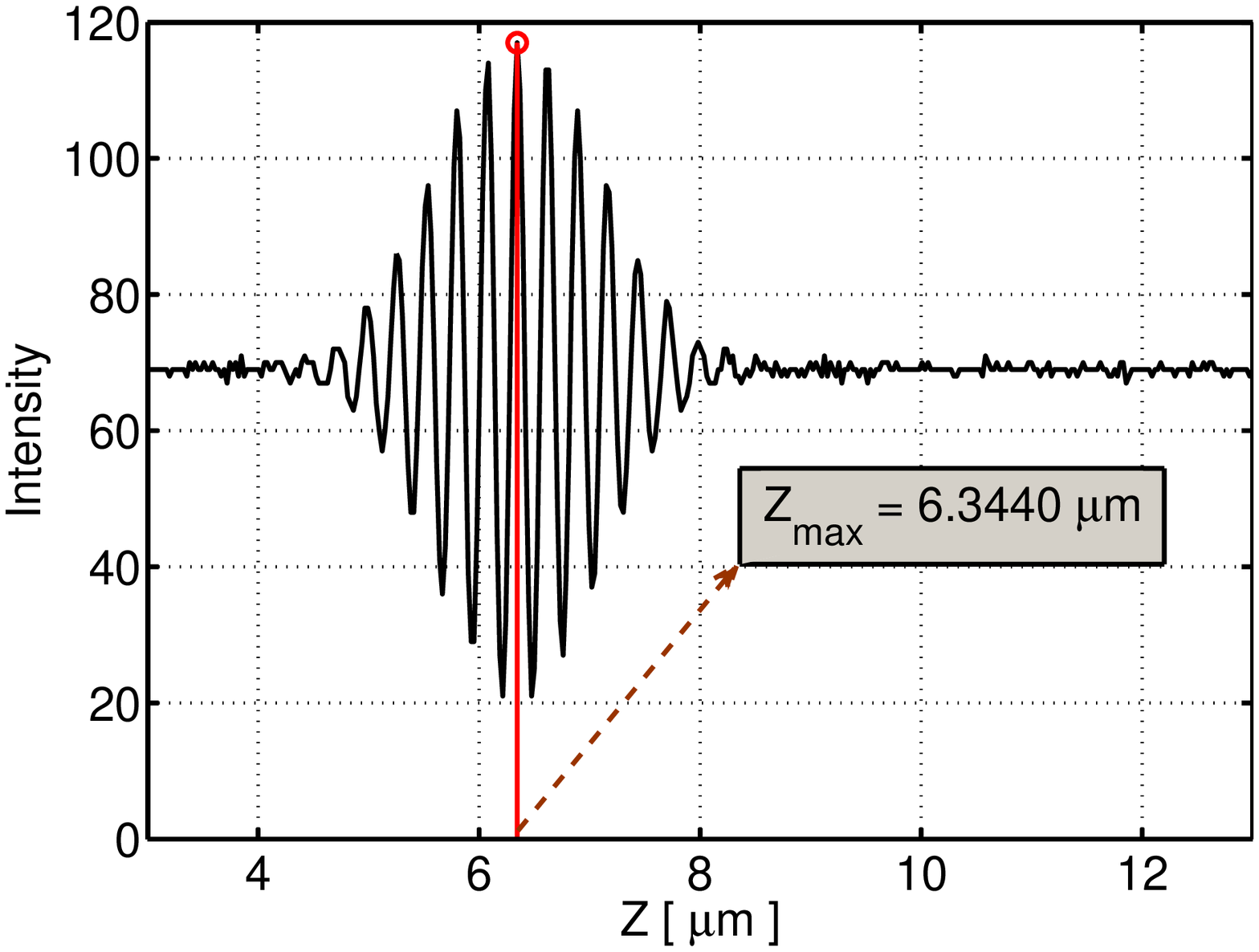}}
\caption{ }
\label{fig:Intensity_high_R}
\end{subfigure}
\qquad%  ~ %add desired spacing between images, e. g. ~, \quad, \qquad, \hfill etc. 
    \begin{subfigure}[b]{0.3\textwidth}
    \captionsetup{justification=centering}
    \fbox{\includegraphics[width=\textwidth]{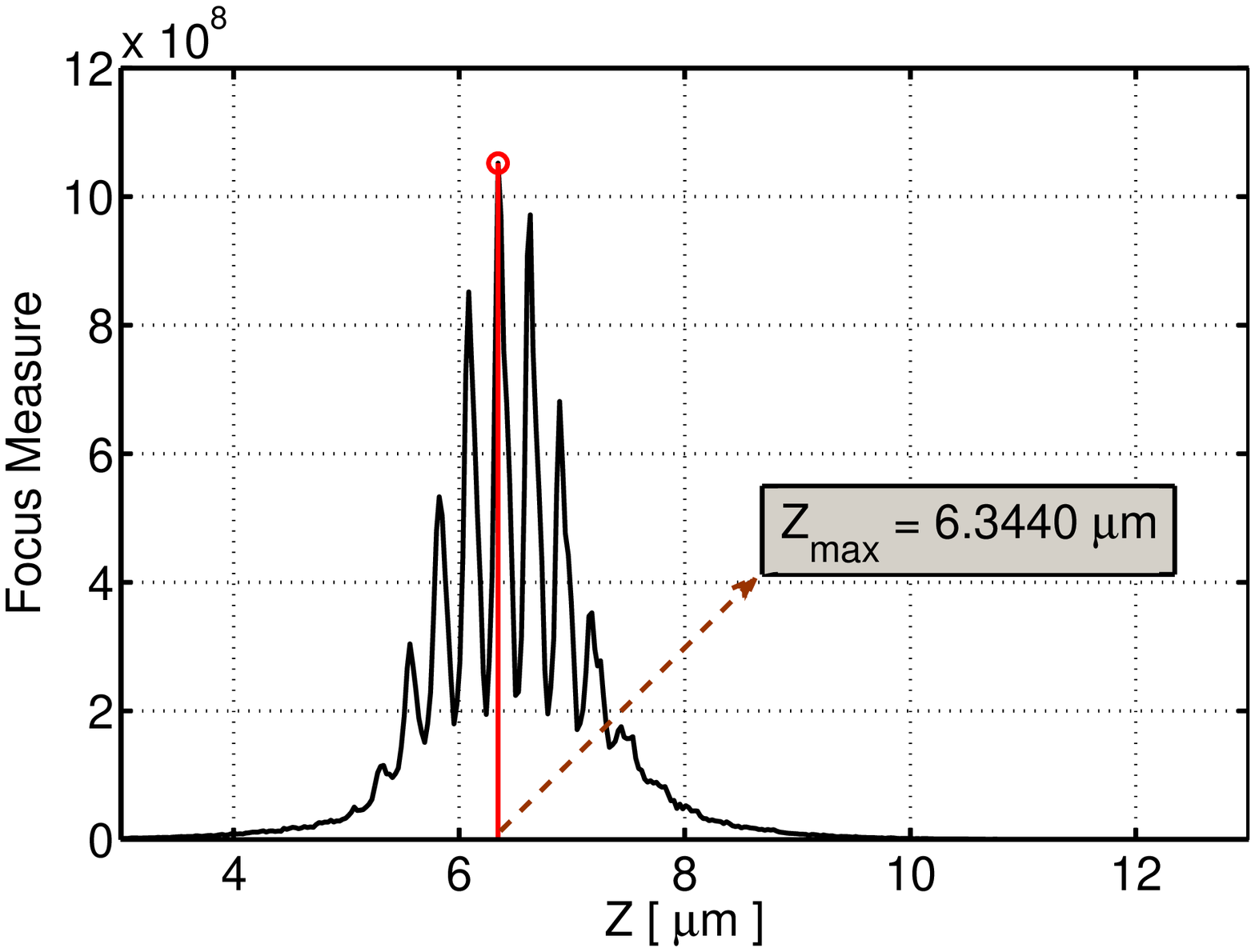}}
        \caption{ }
 \label{fig:FM_high_R}
 \end{subfigure} \\
  % \qquad%  ~ %add desired spacing between images, e. g. ~, \quad, \qquad, \hfill etc. 
    \begin{subfigure}[b]{0.3\textwidth}
    \captionsetup{justification=centering}
    \fbox{\includegraphics[width=\textwidth]{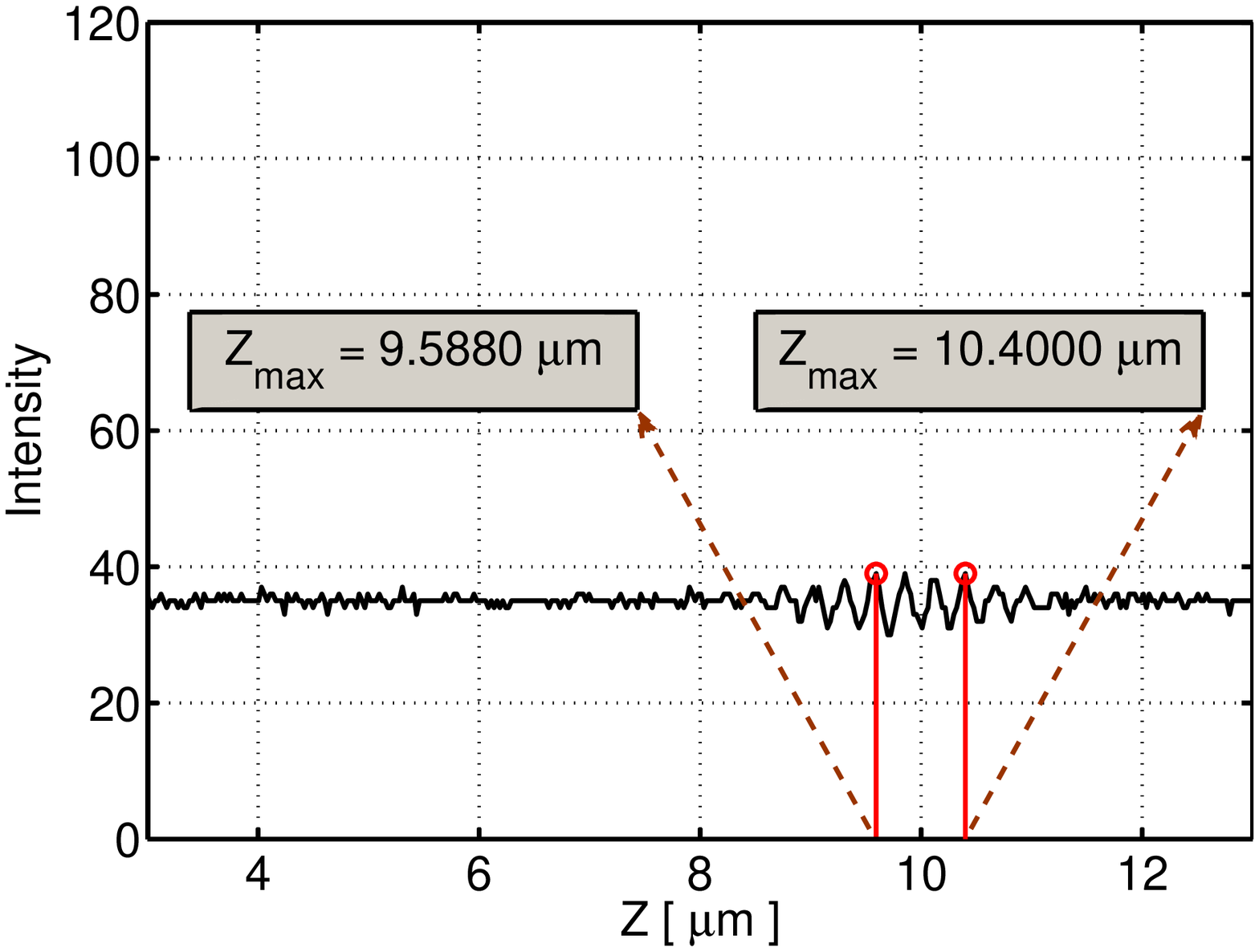}}
 \caption{ }
 \label{fig:Intensity_low_R}
    \end{subfigure}
    \qquad
    \begin{subfigure}[b]{0.3\textwidth}
    \captionsetup{justification=centering}
    \fbox{\includegraphics[width=\textwidth]{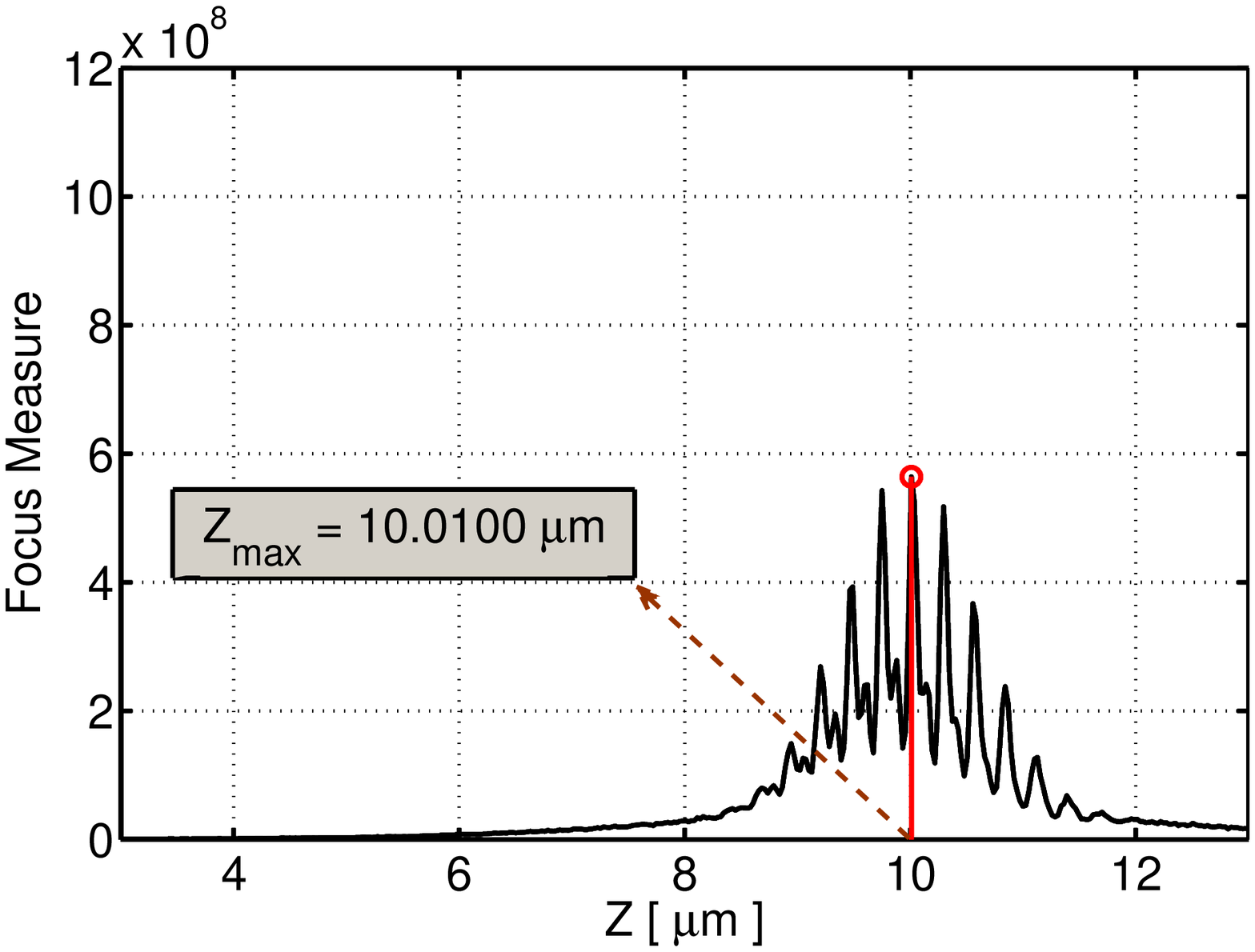}}
    \caption{ }
\label{fig:FM_low_R}
\end{subfigure}
\caption{Comparison of the WLSI intensity criterion against the SFF-WLSI technique for an object with $R_a=49.58$~nm. (a) Intensity signal at a point with reflectance $R = 0.910$ (SNR = 34.08~dB), (b) Focus measurement value at a point with reflectance $R = 0.910$ (SNR = 35.09~dB). (c) Intensity signal at a point with reflectance $R = 0.059$ (SNR = 29.72~dB), and (d) Focus measurement value at a point with reflectance $R = 0.059$ (SNR = 36.18~dB).}
\label{fig:WLSI_and_FM_curves}
\end{figure*}

In the following experiments, we carried out the 3D reconstruction by WLSI via the detection of the maximum of the intensity because we have fine vertical scan stepping and relatively high surface slope~\cite{harding2013handbook}.
To assess the advantage of using focus metrics in WLSI, we have taken two points on the object shown in Fig.~\ref{fig:Interferograms}. One with low reflectance ($R=0.059$), shown with a dashed white line circle,
and another with high reflectance ($R=0.910$), shown with a solid white line circle.
At each of these points, the TENV metric was evaluated using a window $\Omega$ of $5 \times 5$ pixels, and the intensity was calculated for each $Z$ position of the series of images. 
The curves for the high reflectance point are shown in figures~\ref{fig:Intensity_high_R} and~\ref{fig:FM_high_R}. The reference $z$ value is $Z_{\text{ref}}=6.3206~\mu$m. The detected best focus position from the intensity curve (WLSI) is $6.3440~\mu$m and with the focus metric (SFF-WLSI) is $6.3440~\mu$m. The errors are $0.0234~\mu$m, and $0.0234~\mu$m, respectively. Both errors are equal and small ($ < 1 \% $). Also, figures~\ref{fig:Intensity_low_R} and~\ref{fig:FM_low_R} correspond to the point of low reflectance and reference $z$ value $Z_{\text{ref}}=9.9706~\mu$m, with the obtained values from WLSI ($9.5880~\mu$m and $10.4000~\mu$m) and SFF-WLSI ($10.0100~\mu$m), respectively. The errors are $0.3826~\mu$m, $0.4294~\mu$m, and $0.0394~\mu$m; with the problem of not being able to clearly identify a single maximum in Fig.~\ref{fig:Intensity_low_R}.

In Fig.~\ref{fig:Intensity_high_R} we observe that using conventional WLSI the maximum of the intensity signal is determined without difficulty. Likewise, in Fig.~\ref{fig:FM_high_R} we show that by using SFF-WLSI it is also possible to determine $Z_{\max}$ unambiguously. That is, in points of high reflectance WLSI and SFF-WLSI have similar performance. Which makes sense, since both signals have a high signal-to-noise ratio (SNR), 34.08 dB for WLSI and 35.09 dB for SFF-WLSI. In the case of SFF-WLSI, despite the high reflectance of the point, the interference pattern introduces local intensity variations near the point providing enough texture for determining $Z_{\max}$. 

By observing Fig.~\ref{fig:Intensity_low_R} note that the intensity function varies little throughout the $z$-scan and the influence of noise (SNR = 29.72~dB) makes it difficult to determine the position of a single maximum value. In this case, the algorithm finds two maxima. However, for the case of SFF-WLSI, we observe in Fig.~\ref{fig:FM_low_R} that, although the value of the metric decreases with respect to the point of high reflectance, the curve allows an accurate determination of $Z_{\max}$ with SNR=36.18~dB. That is, in points of low reflectance the focus technique has a better performance than the maximum intensity detection technique in WLSI.
The previous results show that using a focus metric on WLSI images (SFF-WLSI) allows determining the focus position robustly under the criterion of the maximum value of the metric in both points of high and low reflectance.

\subsection{3D Reconstruction Performance Assessment by Varying Surface Roughness}
\label{sec:DesWLSI_FMWLSI}

In the previous sections, the advantage of using SFF-WLSI over SFF and  WLSI was shown. Now, we analyze a global performance of such techniques by performing 3D reconstruction of the surface of objects with different roughness. We simulated ten objects of Gaussian shape with different roughness calculated by~\eqref{eq:reflectivity}. 
The arithmetic roughness ($R_a$) is in the range $49.58~\text{nm} \geq R_a \geq 34.47~\text{nm}$. We have taken this roughness range because it is above the resolution of the WLSI technique and below the resolution of the SFF technique.

The performance of each technique was evaluated by measuring the RMSE of the reconstructed surface with respect to the reference surface. In table~\ref{tab:RMSE_table} we show the obtained results. Note that the WLSI technique has the worst performance on surfaces of high roughness. We also observe that although the SFF technique has an adequate performance on surfaces with high roughness, it has poor performance on surfaces with low roughness. However, the proposed SFF-WLSI method works effectively on both high and low roughness surfaces. 

\begin{figure*}[htbp]
\centering
\fbox{\begin{minipage}[t]{0.3\linewidth}
 \centering
 SFF\\
 %\noindent\makebox[\linewidth]{\rule{\linewidth}{0.5pt}}
 \makebox[0pt][c]{\rule[3pt]{5.7cm}{1pt}}\\
% \underline{\hspace{ \textwidth}} \\
 %\noindent\rule{0.8\textwidth}{1pt}\\
%\hspace{0.02\textwidth}
 \begin{subfigure}[b]{0.8\textwidth}
\captionsetup{justification=centering}
\fbox{\includegraphics[width=\textwidth]{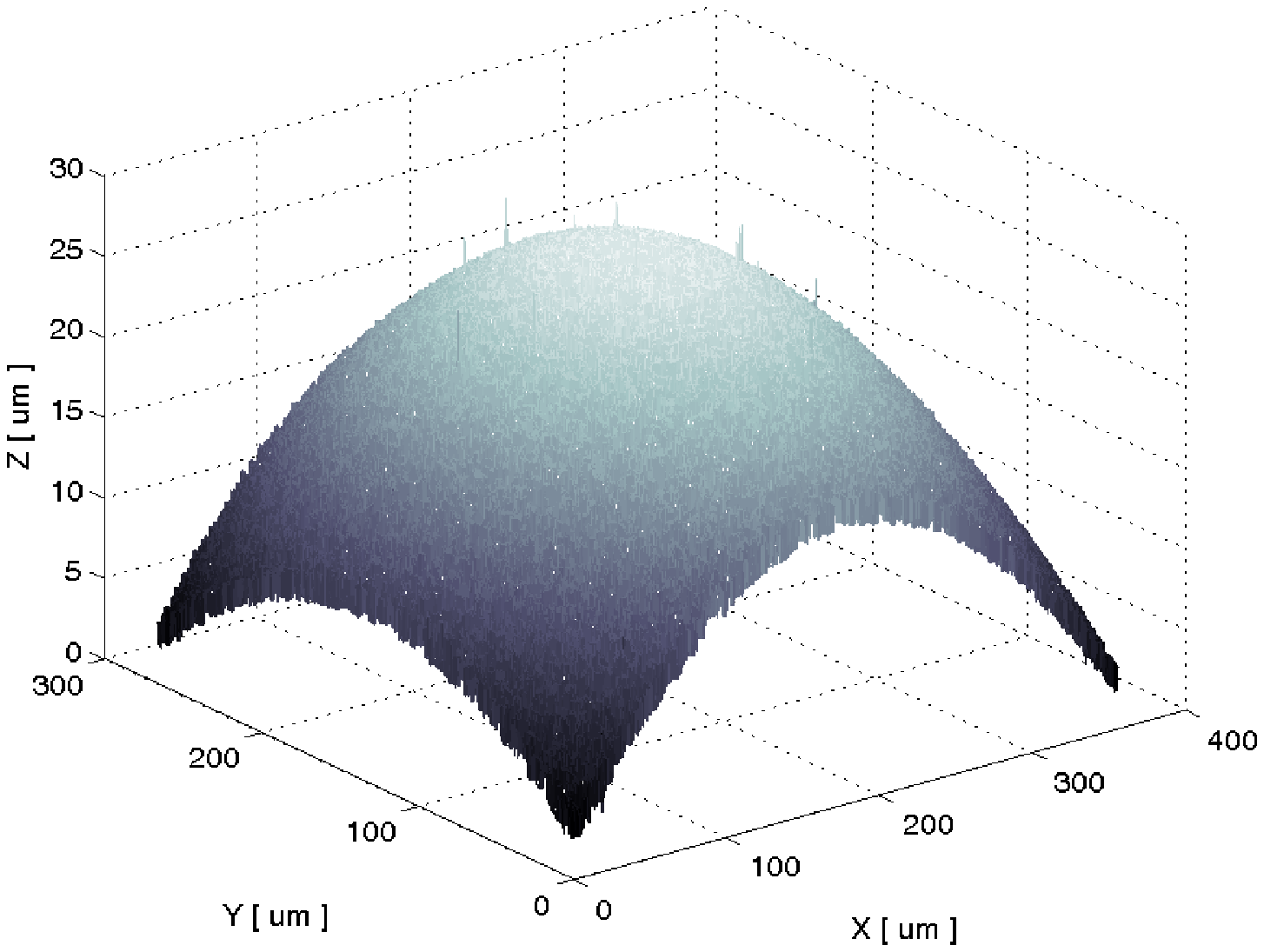}}
\caption{ }
\label{fig:3DFM_high_R}
\end{subfigure}\\
\begin{subfigure}[b]{0.8\textwidth}
    \captionsetup{justification=centering}
    \fbox{\includegraphics[width=\textwidth]{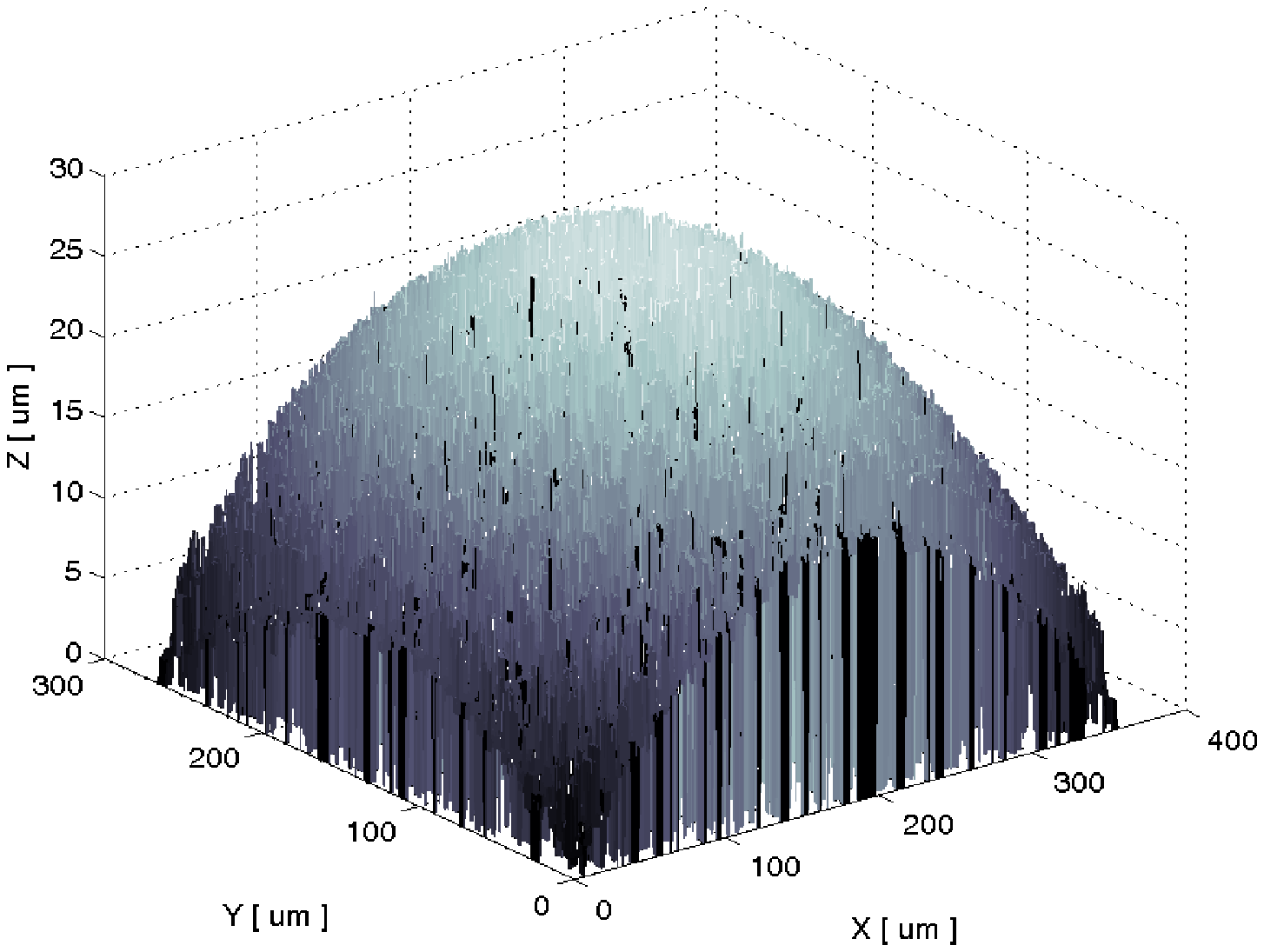}}
    \caption{ }
\label{fig:3DFM_low_R}
\end{subfigure}
\end{minipage}}
\fbox{ \begin{minipage}[t]{0.3\linewidth}
 \centering
  WLSI\\
  \makebox[0pt][c]{\rule[3pt]{5.7cm}{1pt}}\\
    \begin{subfigure}[b]{0.8\textwidth}
    \captionsetup{justification=centering}
    \fbox{\includegraphics[width=\textwidth]{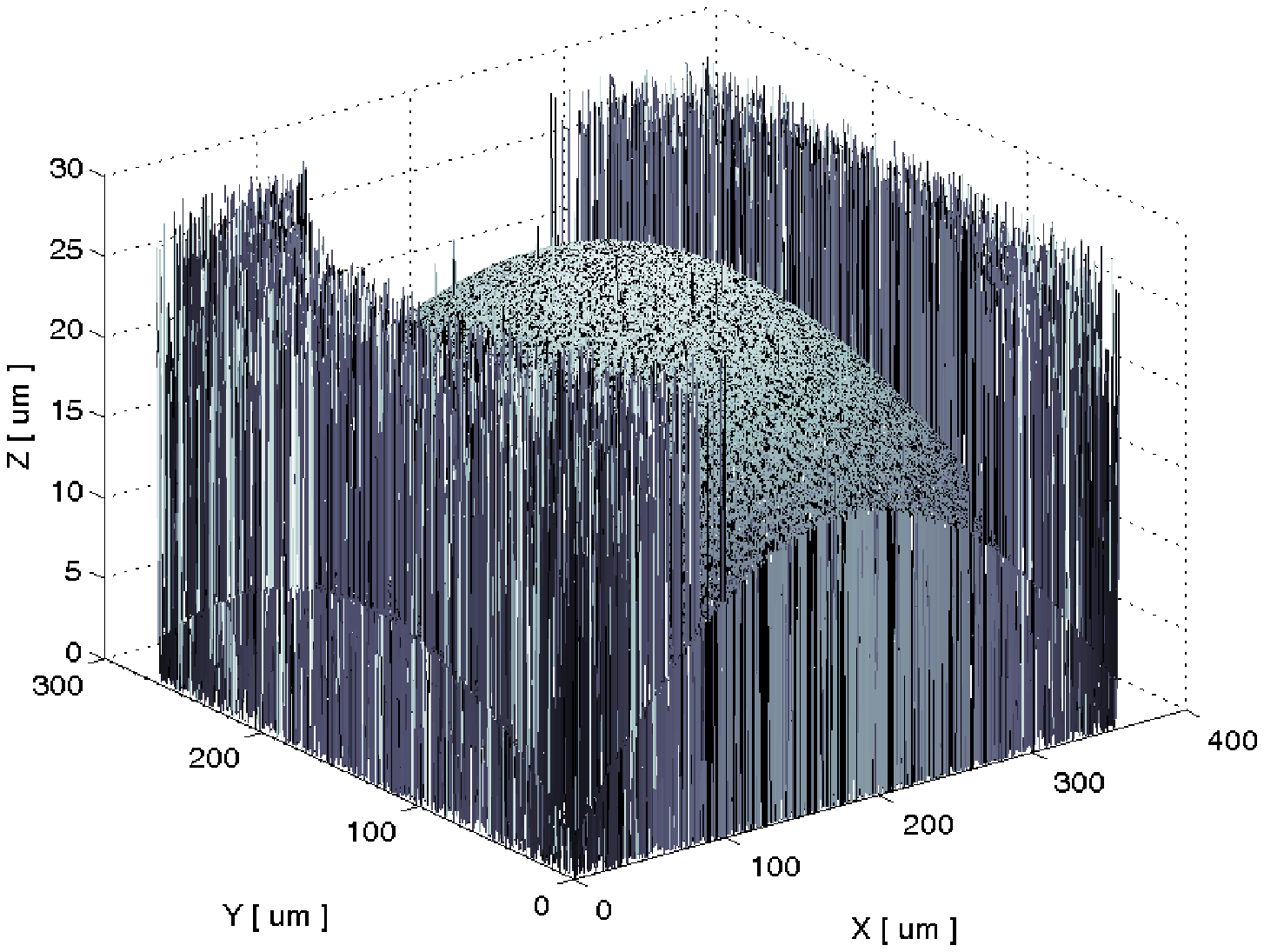}}
        \caption{ }
         \label{fig:3DWLSI_high_R}
 %\label{fig:3DFM_low_R}
 \end{subfigure}\\
 \begin{subfigure}[b]{0.8\textwidth}
    \captionsetup{justification=centering}
    \fbox{\includegraphics[width=\textwidth]{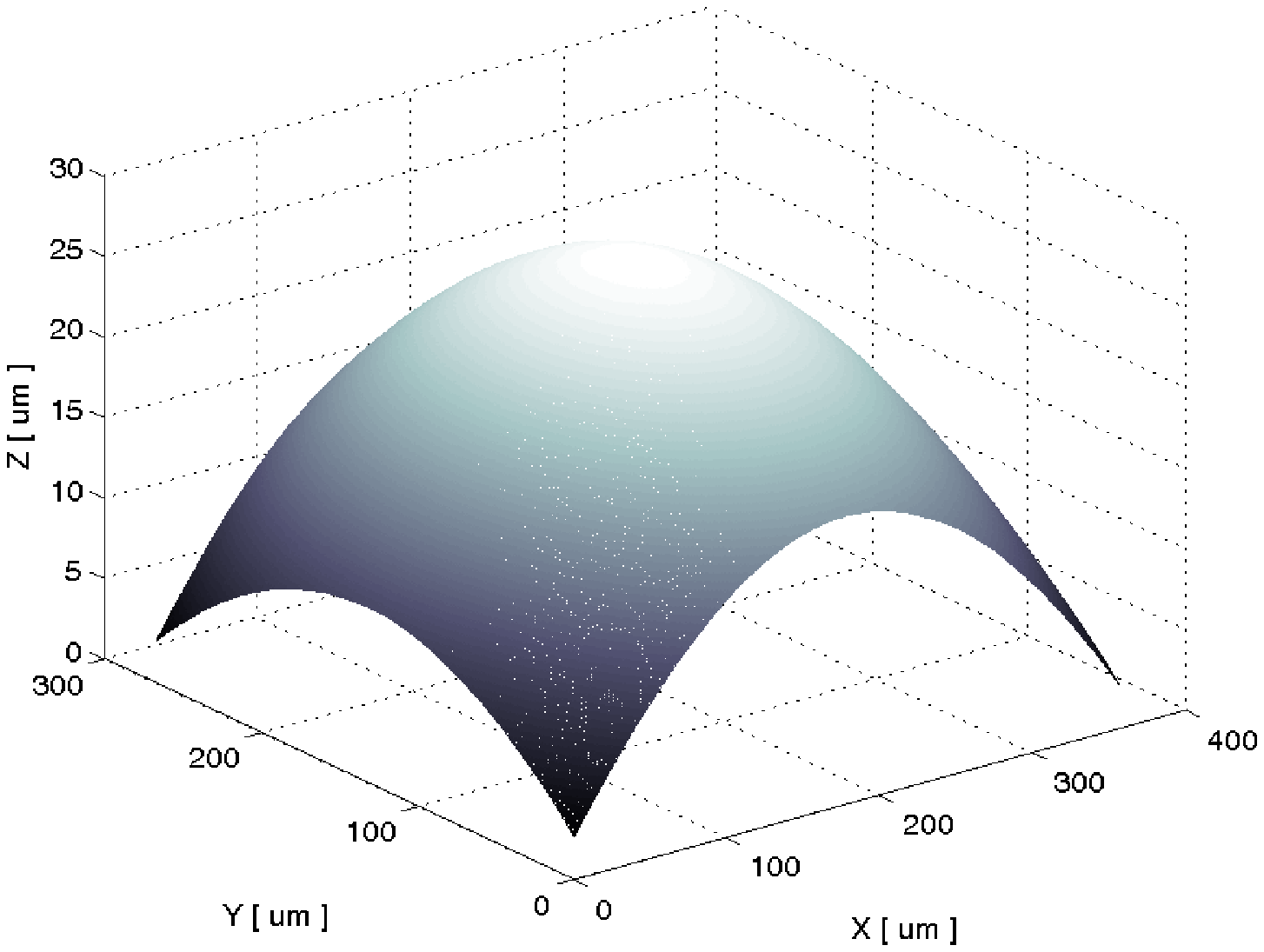}}
        \caption{ }
 \label{fig:3DWLSI_low_R}
    \end{subfigure}
 \end{minipage}}
%\hspace{0.02\textwidth}
\fbox{\begin{minipage}[t]{0.3\linewidth}
 \centering
 SFF-WLSI\\
\makebox[0pt][c]{\rule[3pt]{5.7cm}{1pt}}\\
\begin{subfigure}[b]{0.8\textwidth}
\captionsetup{justification=centering}
\fbox{\includegraphics[width=\textwidth]{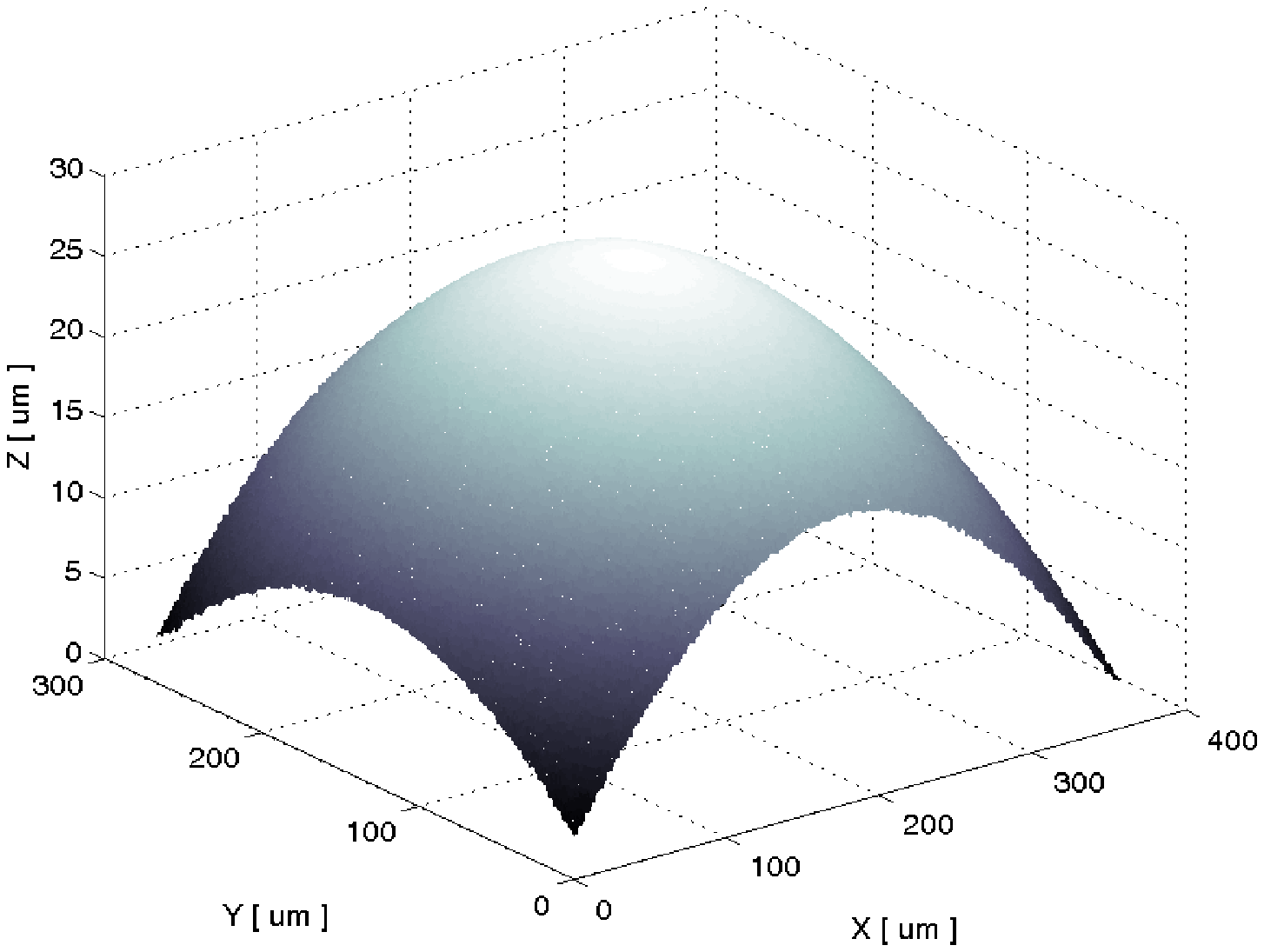}}
\caption{ }
\label{fig:3DFM_WLSI_high_R}
\end{subfigure}\\
\begin{subfigure}[b]{0.8\textwidth}
\captionsetup{justification=centering}
\fbox{\includegraphics[width=\textwidth]{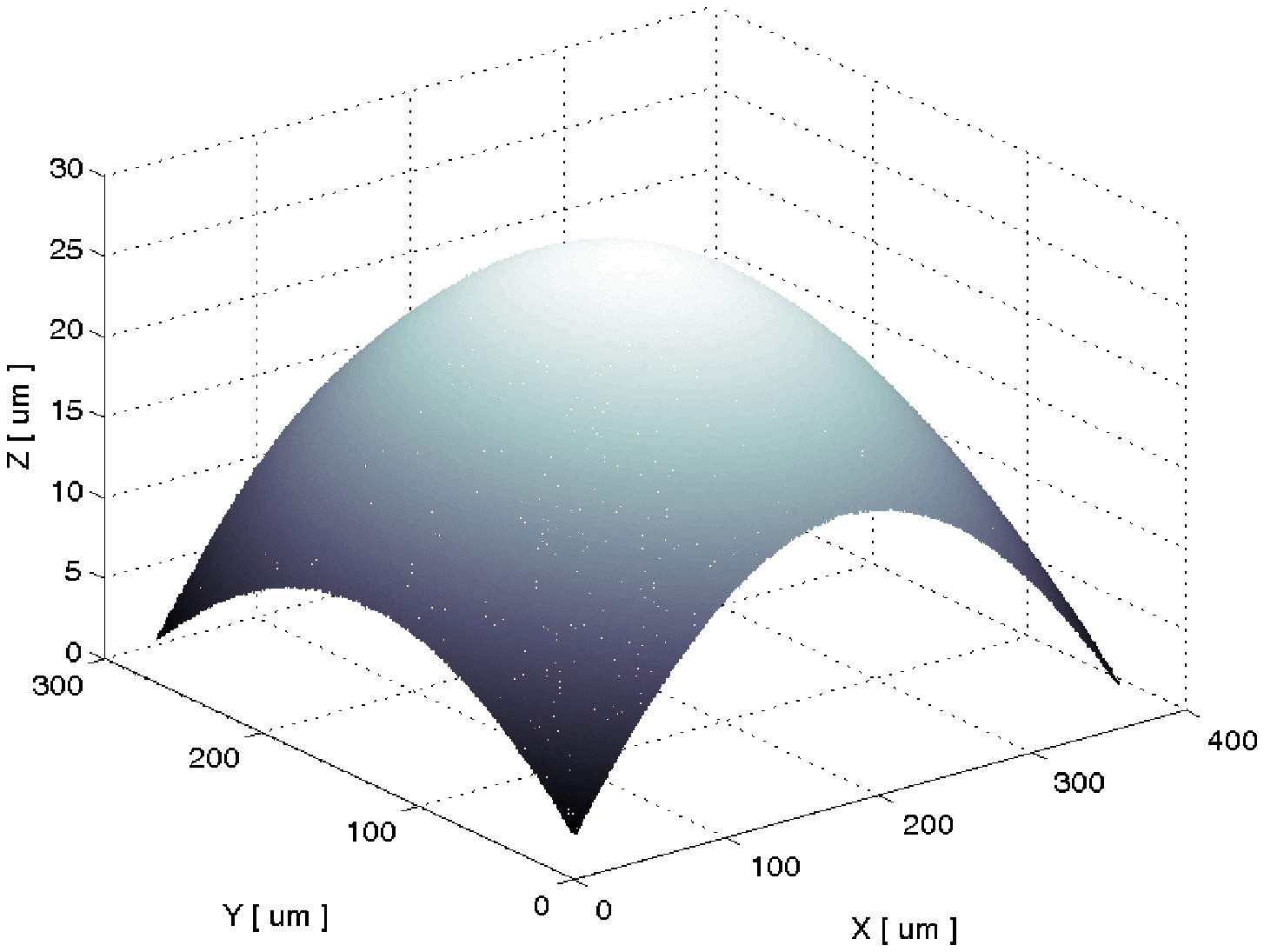}}
\caption{ }
\label{fig:3DFMWLSI_low_R}
\end{subfigure} 
\end{minipage}}
\caption{3D Reconstruction of a Gaussian-shaped object. Top row: roughness $R_a=49.58$~nm. Bottom row: roughness $R_a=34.47$~nm. (a) and (b) SFF in standard microscopy. (c) and (d) WLSI. (e) and (f) proposed SFF-WLSI method.}
\label{fig:3D_topography}
\end{figure*}

In Fig.~\ref{fig:3D_topography} we show the 3D reconstructions of the objects of highest and lowest roughness from table~\ref{tab:RMSE_table}. The top row corresponds to the surface of high roughness and the bottom row to the surface of low roughness.  From left to right, each column corresponds to the 3D reconstruction by SFF, WLSI, and SFF-WLSI, respectively. From the 3D reconstructions and the RMSE values from table~\ref{tab:RMSE_table}, the proposed method performs better than SFF or WLSI alone in high roughness samples (Maps \#1 to \#5), whereas, in low roughness (Maps \#6 to \#10), SFF-WLSI and WLSI have comparable performance. Moreover, SFF under-performs in these samples. 

% \newpage

\section{Experimental results in real Objects}
\label{sec:resul_real_object}

We used an Olympus BH microscope equipped with a motorized positioning stage, a $10 \times$ Mirau type interference objective lens and a Tucsen IS 130 CCD camera. 
We acquired a WLSI image series with the stage displaced $0.018~\mu$m for each frame.

\begin{figure}[t]
\centering
\begin{subfigure}[b]{0.38\textwidth}
\captionsetup{justification=centering}
\fbox{\includegraphics[width=\textwidth]{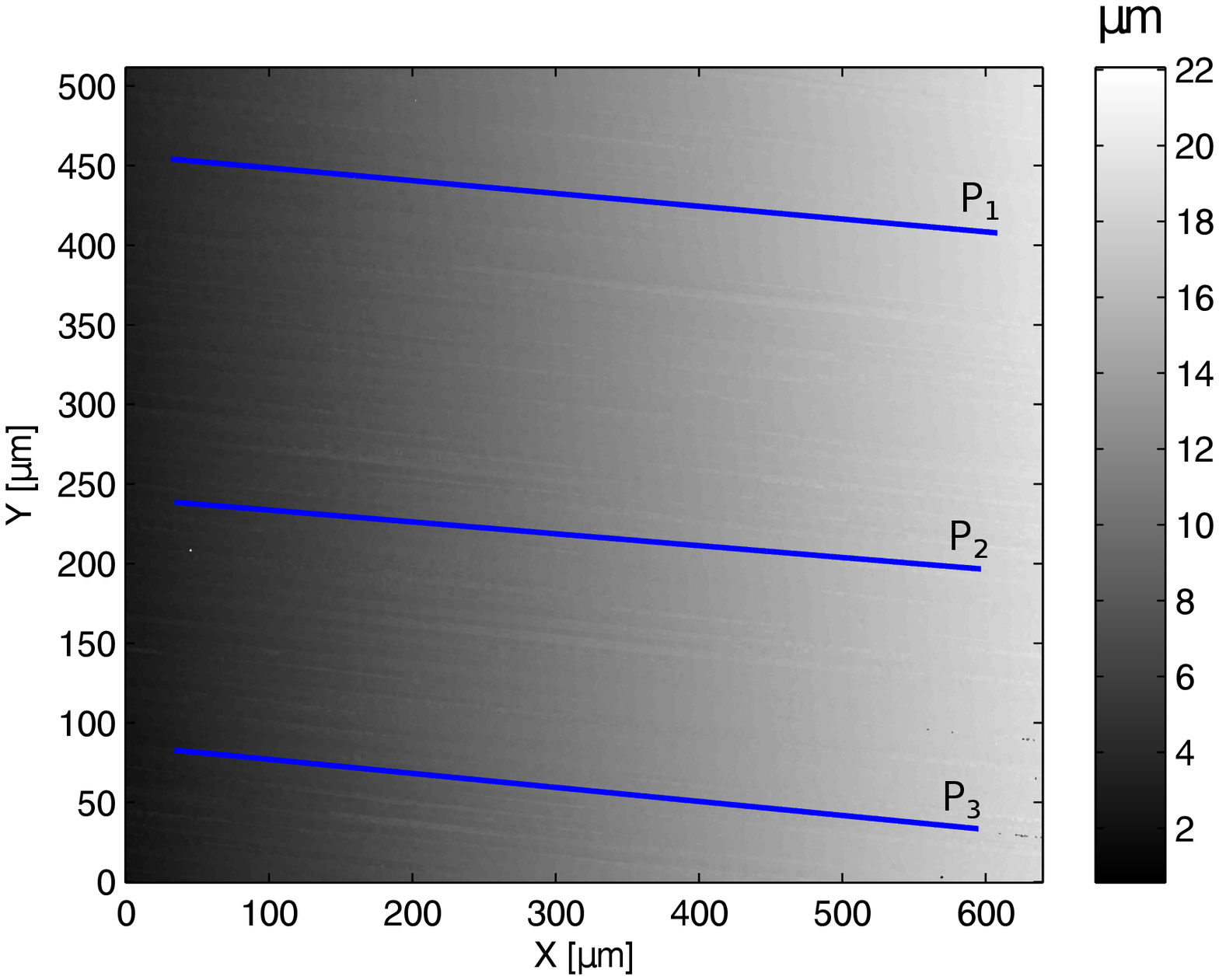}}
\caption{ }
\label{fig:3Dstandarmap}
\end{subfigure}
\qquad
    \begin{subfigure}[b]{0.38\textwidth}
    \captionsetup{justification=centering}
    \fbox{\includegraphics[width=\textwidth]{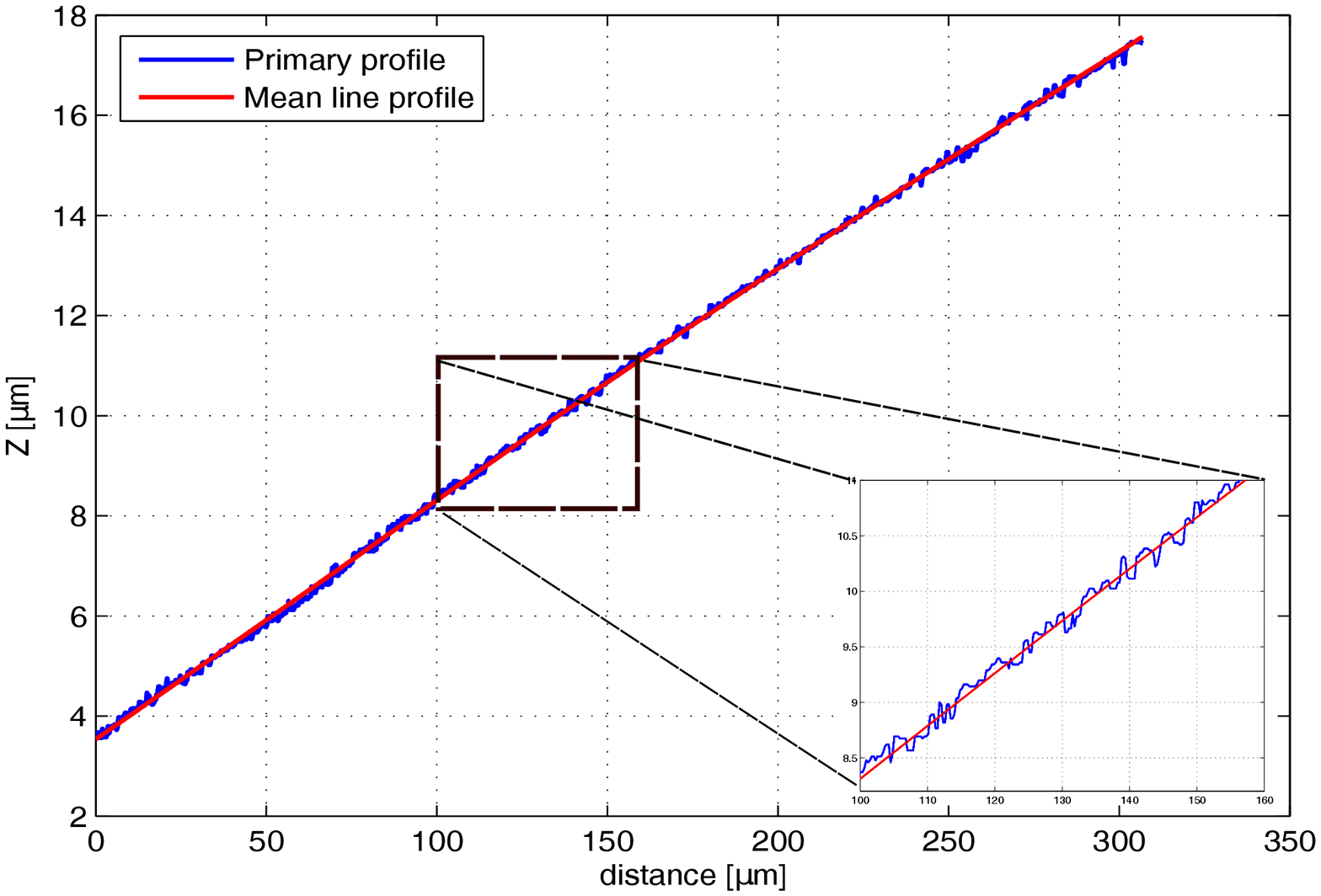}}
        \caption{ }
         \label{fig:standarProfil}
 %\label{fig:3DFM_low_R}
 \end{subfigure}
\caption{3D reconstruction of a flat lapping specimen with SFF-WLSI. (a) Reconstructed surface map, and (b) a profile and its mean line profile along the $P_1$ direction shown in (a).}
\label{fig:Standard_Result}
\end{figure}

To validate the proposed method, we used the flat lapping specimen of a Rubert \& Co.~Ltd roughness comparison set No.~130 with roughness parameter $R_a=0.05~\mu$m. We reconstructed a portion of the specimen as shown in Fig~\ref{fig:3Dstandarmap}. In Fig.~\ref{fig:standarProfil} we show a profile and its mean line profile along the $P_1$ direction.
We estimated the mean line profile according to the standard  ISO/TS 16610-22, and the $R_a$ roughness parameter was calculated by following the standard EN ISO 4287. We obtained $R_a$ values  of $0.046~\mu$m, $0.057~\mu$m, and $0.063~\mu$m through the directions $P_1$, $P_2$, and $P_3$ respectively, as shown in Fig.~\ref{fig:3Dstandarmap}. The average value of $R_a = 0.055~\mu$m and standard deviation $\sigma=0.008~\mu$m, which is quite close to the reference value of $R_a=0.05~\mu$m.
We also performed the 3D reconstruction via WLSI with maximum intensity detection yielding an average value of $R_a=0.047~\mu$m and standard deviation $\sigma=0.008~\mu$m. As expected, in this sample of high reflectivity the measurements with both methods agree within the experimental error.

\begin{figure}[t]
\centering
\begin{subfigure}[b]{0.35\textwidth}
\captionsetup{justification=centering}
\fbox{\includegraphics[width=\textwidth]{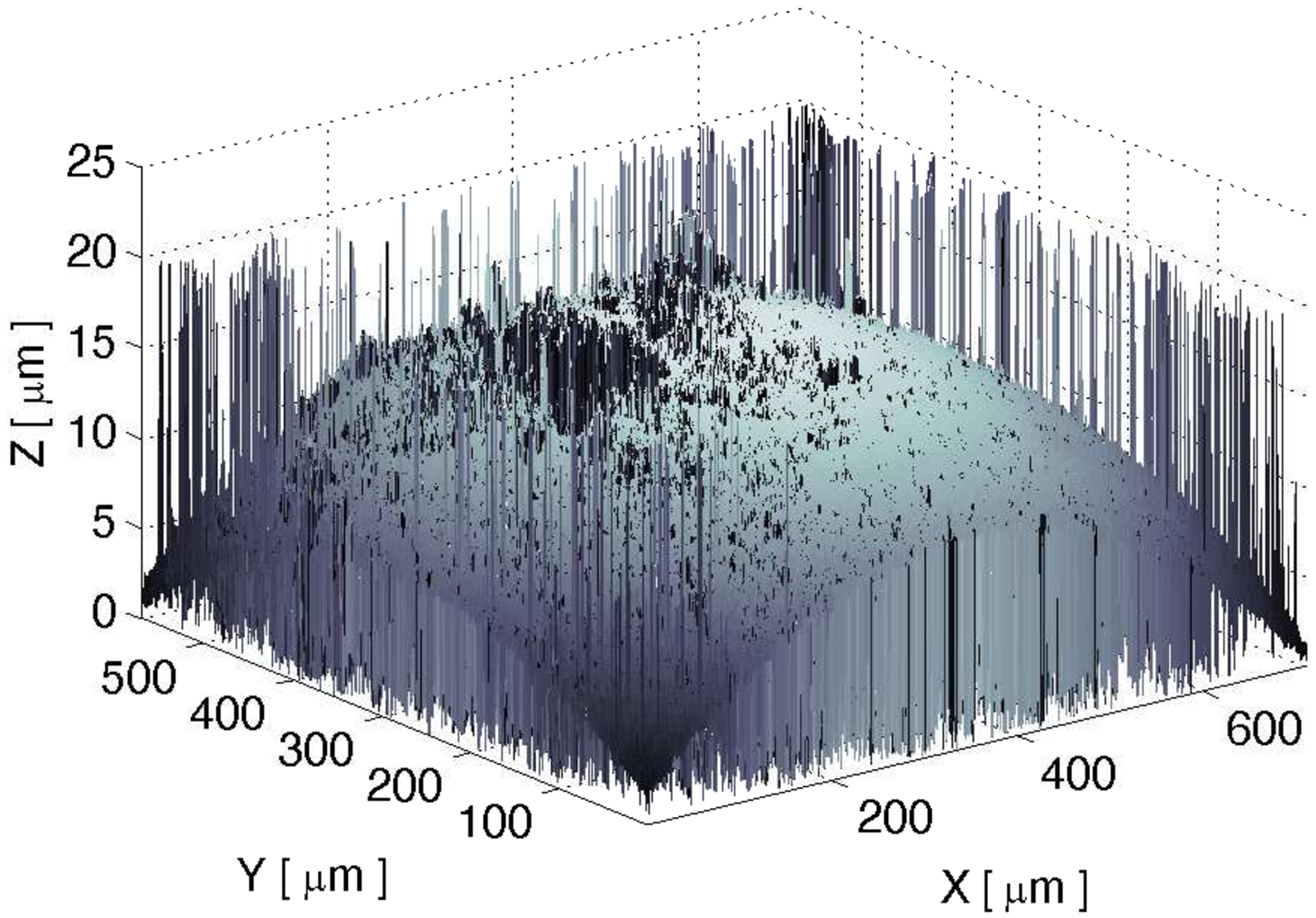}}
\caption{ }
\label{fig:3D_WLSI_expe}
\end{subfigure}
\qquad
    \begin{subfigure}[b]{0.35\textwidth}
    \captionsetup{justification=centering}
    \fbox{\includegraphics[width=\textwidth]{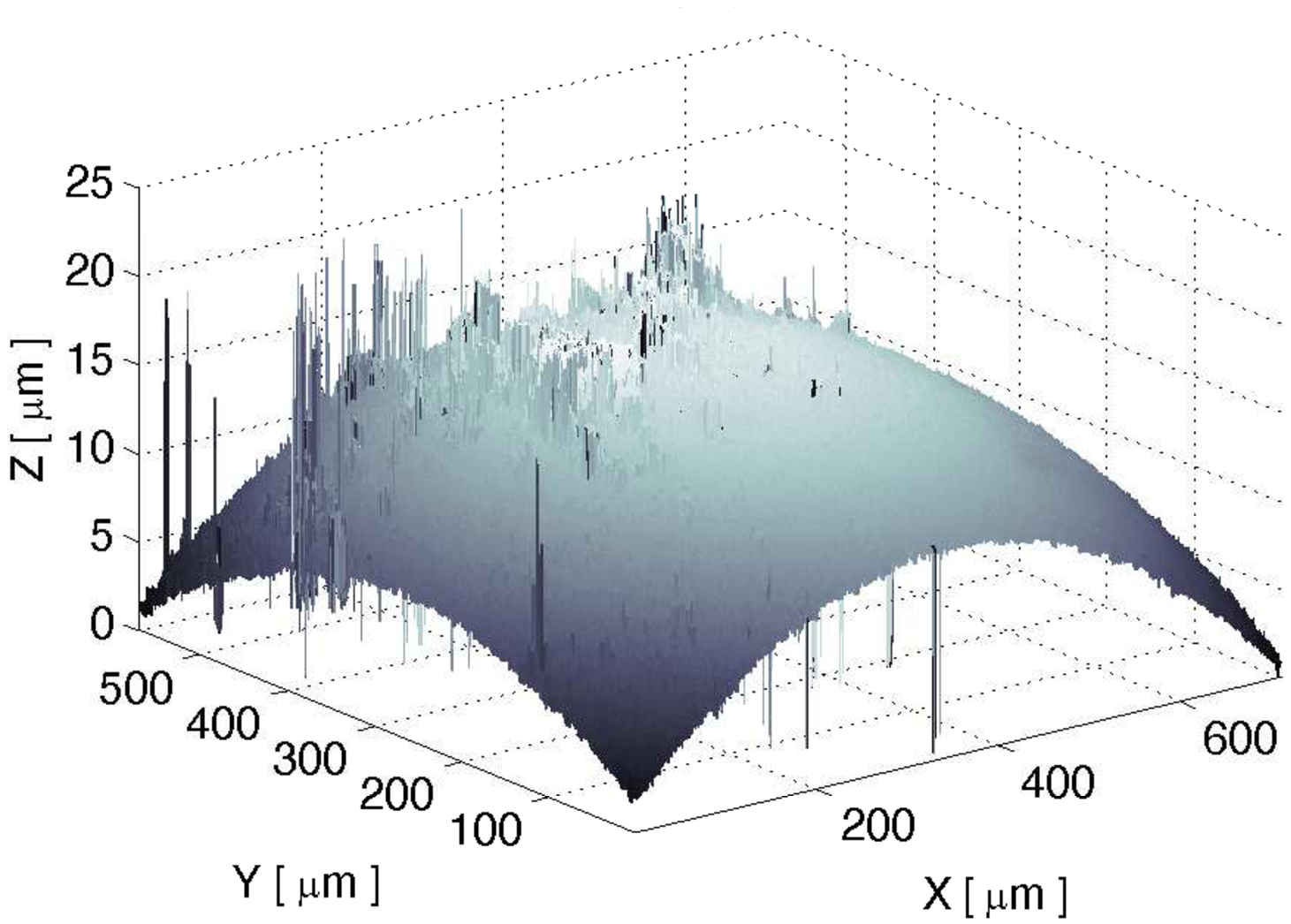}}
        \caption{ }
         \label{fig:3D_FM_WLSI_expe}
 %\label{fig:3DFM_low_R}
 \end{subfigure}
\caption{A metallic sphere topography reconstructed with: (a)~WLSI, and (b) SFF-WLSI.}
\label{fig:3D_Metallic_sphere}
\end{figure}

\begin{figure}[t]
\centering
% \captionsetup{justification=centering}
\fbox{\includegraphics[width=\linewidth]{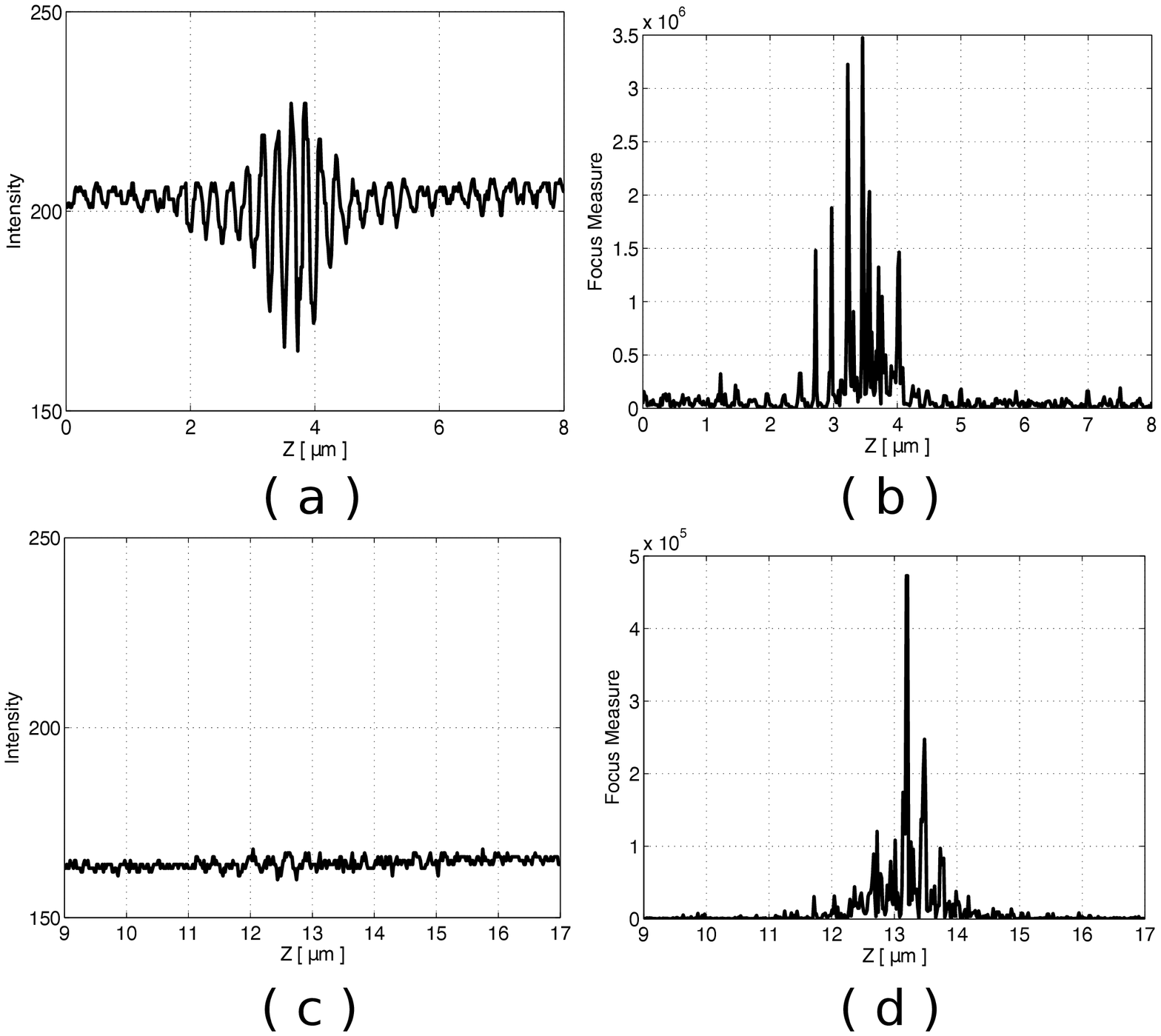}}
\caption{Comparison of WLSI vs SFF-WLSI signals from the metallic sphere in Fig.~\ref{fig:3D_Metallic_sphere}. (a) WLSI and (b) SFF-WLSI signals for a high reflectivity point. (c) WLSI and (d) SFF-WLSI signals for a low reflectivity point.}
\label{fig:3D_Metallic_sphere_curves}
\end{figure}

The second reconstructed real object was the surface of a metal sphere in the process of corrosion. We obtained the topography of the sample by WLSI and SFF-WLSI. We show the 3D reconstructions in Fig.~\ref{fig:3D_Metallic_sphere} which have not been post-processed. We observe that the 3D reconstruction by SFF-WLSI (Fig.~\ref{fig:3D_FM_WLSI_expe}) yields a robust measurement of the surface topography with less unresolved height values. In Fig.~\ref{fig:3D_Metallic_sphere_curves} we show the comparison of the WLSI versus SFF-WLSI signals for a point of high reflectivity and low reflectivity. Note that the resulting curves are in agreement with the results from the simulation shown in Fig.~\ref{fig:WLSI_and_FM_curves}. The SFF-WLSI response for the low reflectivity point with a unique maximum and high SNR is clearly an improvement over the WLSI signal.
These results show how the low reflectivity caused by corrosion hinders the WLSI measurement. 
% We have also included two 3D surface reconstructions with the proposed method as supplementary material. A part of a coin (Visualization 4) and a vertical milling specimen (Visualization~5).

\section{Conclusions}
\label{sec:conclusion}

Accurate surface metrology requires robust measurement techniques that return reliable results under variable surface conditions such as reflectivity and high roughness. We carried out simulations of two profilometers with different measurement techniques, namely Shape from Focus (SFF), White Light Scanning Interferometry (WLSI), and the proposed SFF-WLSI with the TENV focus metric, along with different surface roughness. Through these simulations, we reproduced conditions for performance comparisons which cannot be carried out in practice. 

We showed that processing a WLSI stack of images with a focus metric (SFF-WLSI), such as the TENV metric, yielded accurate surface measurements under different surface roughness and surface reflectivity outperforming the conventional WLSI and the SFF techniques. This robustness is due to the capacity of the SFF-WLSI technique to determine the position of maximum focus in low reflectivity regions because it takes into account information from neighboring points. In high reflectivity regions where there is hardly any texture from the object, the interference pattern introduces local texture which improves the determination of the best focus position. We validated the simulation results on two real objects. The first, a flat lapping specimen with $R_a=0.05~\mu$m for which we measured an average value of $R_a ~=~0.055 ~\mu$m and standard deviation $\sigma=0.008~\mu$m. The second, a metallic sphere with corrosion for which the proposed SFF-WLSI method produced a better 3D reconstruction with less undefined depth values.
 % we reconstructed with WLSI versus the proposed SFF-WLSI technique producing a

\section*{Acknowledgment}
This work has been partly funded by Colciencias project 538871552485 and Colciencias doctoral support program 785-2017.
Parts of this work were presented at the Imaging and Applied Optics Congress 2018, Orlando, Florida, 2018, p. JTu4A.19.
% % Bibliography
\bibliography{sample}
\bibliographystyle{osajnlnt}

\section*{APPENDIX}

\begin{figure}[tbp]
\centering
\fbox{\includegraphics[width=0.8\linewidth]{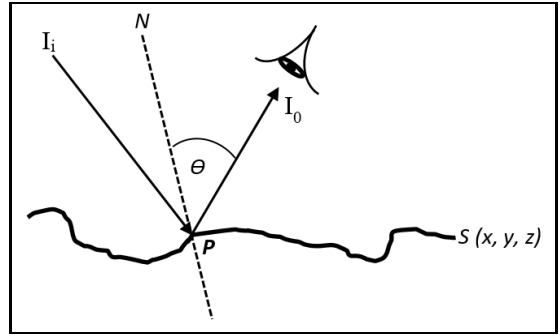}}
\caption{Geometry for Lambert’s cosine law for intensity.}
\label{fig:Lambert_law}
\end{figure}

To determine the reflected intensity $I_0$ at a point $P$ on a surface using the Lambert's model~\eqref{eq:Lamb_law}, we must determine $ \cos \theta $, which will depend on the normal $ N $ at each point on the surface, as illustrated in Fig.~\ref{fig:Lambert_law}.
Knowing the analytic expression of a surface of level $S (x, y, z) = C$, we can find the direction of the normal $N$ through the direction of the gradient vector of the surface by
\begin{equation}
\vec{N}=\vec{\bigtriangledown} S.
\label{eq:grad}
\end{equation}
For the observation from the $+z$ direction, from the scalar product we obtain
\begin{equation}
\vec{\bigtriangledown} S\cdot \widehat{k}=\left | \vec{\bigtriangledown} S \right | \cos \theta. 
\label{eq:grad2}
\end{equation}
Therefore,
\begin{equation}
\cos \theta= \frac{  \vec{\bigtriangledown} S \cdot \widehat{k} }{\left | \vec{\bigtriangledown} S \right |} 
\label{eq:cos}
\end{equation}
Assuming the surface as the sum of a surface $\mathcal{W}$ and a function of roughness $R$,
\begin{equation}
z(x,y)=\mathcal{W}(x,y)+R(x,y). 
\label{eq:supZ}
\end{equation}
Thus, the level surface is
\begin{equation}
S (x, y, z) = z-\mathcal{W} (x, y) -R (x, y) = 0, 
\label{eq:level_surf}
\end{equation}
and the gradient is expressed as
\begin{equation}
 \vec{\bigtriangledown} S=-\frac{\partial }{\partial x}\left ( \mathcal{W}+R \right )\widehat{i}-\frac{\partial }{\partial y}\left ( \mathcal{W}+R \right )\widehat{j}+\frac{\partial }{\partial z}\left ( z \right )\widehat{k}, 
\label{eq:gradS1}
\end{equation}
\begin{equation}
\vec{\bigtriangledown} S = -\left ( \frac{\partial\mathcal{W}}{\partial x} +\frac{\partial R}{\partial x}\right )\widehat{i}-\left ( \frac{\partial \mathcal{W}}{\partial y} +\frac{\partial R}{\partial y}\right )\widehat{j}+\widehat{k}, 
\label{eq:gradS2}
\end{equation}
Then, \eqref{eq:cos} can be written as
\begin{equation}
\cos \theta= \frac{1}{\sqrt{\left ( \frac{\partial \mathcal{W}}{\partial x} +\frac{\partial R}{\partial x}\right )^2+\left ( \frac{\partial \mathcal{W}}{\partial y} +\frac{\partial R}{\partial y}\right )^2+1}}. 
\label{eq:cos2}
\end{equation}

To model the roughness we used the multivariate Weierstrass-Mandelbrot (W-M)  developed by Ausloos and Berman given by~\eqref{eq:reflectivity}. 

% The multivariate W-M function in Cartesian coordinates is expressed as:
% % 
% \begin{equation}
% \begin{split}
% R(x,y) =  &C\sum\limits_{m=1}^M\sum\limits_{n=-\infty}^\infty \gamma^{(D_s-3)n} \times    
% \Bigg \{
% \cos\Phi_{m,n}~- \\ &\cos \left[ \frac{2\pi \gamma^n (x^2+y^2)^{\frac{1}{2}}}{L} \cos\left( \arctan \left(\frac{y}{x} \right) - \frac{\pi m}{M} \right) + \Phi_{m,n} \right]
% \Bigg \} \enspace, 
% \end{split}
% \label{eq:WM_equa}
% \end{equation}
% %    
% where $C = L(\frac{K}{L})^{D_s-2} \left ( \frac{\ln \gamma}{M} \right )^\frac{1}{2}$, $D_s$ is the fractal dimension, $2 <D_s <3$ and $\gamma> 1$. $K$ is a constant that acts as a scale factor, $L$ the maximum extension of the surface to be simulated, $M$ is the number of modes for the roughness simulation and $\Phi_{m, n}$ a random phase in the range $[0,2\pi]$.
% 
Here, we take the example for a Gaussian surface
\begin{equation}
\mathcal{W}(x,y)=A\exp \left [ -\alpha (x^2 + y^2) \right ]  \enspace,
\label{eq:Gauss_surface}
\end{equation}
where A and $ \alpha> 0$ are constants.
Then, by developing the partial derivatives of $G$ and $R$ and simplifying with trigonometric identities, we get
\begin{equation}
\frac{\partial \mathcal{W}}{\partial x}=-2\alpha xA \exp \left [ -\alpha\left ( x^2+y^2 \right ) \right ]  \enspace,
\label{eq:Dx_surface}
\end{equation}
\begin{equation}
\frac{\partial \mathcal{W}}{\partial y}=-2\alpha yA \exp \left [ -\alpha\left ( x^2+y^2 \right ) \right ]  \enspace,
\label{eq:Dy_surface}
\end{equation}
\begin{equation}
%\begin{eqnarray}
\begin{split}
\frac{\partial R}{\partial x}=&\frac{2\pi C}{\mathcal{L}}\sum_{m=1}^{M} \sum_{n=-\infty }^{\infty} \gamma^{(D_s-2)n} \times \cos \left( \frac{\pi m}{M} \right ) \times \\ &\sin \Bigg \{ \frac{2\pi \gamma^n \left ( x^2+y^2 \right )^{1/2}}{\mathcal{L}} \times   \\ & \qquad \; \; \; \;  \cos \left( \arctan \left ( \frac{y}{x} \right ) -  \frac{\pi m}{M}\right ) + \Phi_{m,n}   \Bigg \},
\end{split}
\label{eq:Dx_Roughness}
%\end{eqnarray}
\end{equation}
\begin{equation}
\begin{split}
\frac{\partial R}{\partial y}=&\frac{2\pi C}{\mathcal{L}}\sum_{m=1}^{M} \sum_{n=-\infty }^{\infty} \gamma^{(D_s-2)n} \times \sin \left ( \frac{\pi m}{M} \right ) \times \\ &\sin \Bigg \{ \frac{2\pi \gamma^n \left ( x^2+y^2 \right )^{1/2}}{\mathcal{L}}  \times   \\ & \qquad \; \; \; \;  \cos \left( \arctan \left ( \frac{y}{x} \right ) -  \frac{\pi m}{M}\right ) + \Phi_{m,n}   \Bigg \}.
\end{split}
\label{eq:Dy_Roughness}
\end{equation}
Substituting \eqref{eq:Dx_surface}-\eqref{eq:Dy_Roughness} in \eqref{eq:cos2} and using \eqref{eq:Lamb_law}, the reflected intensity is established for a Gaussian surface with roughness.

\end{document}